\documentclass[aps,prb,10pt,twocolumn,notitlepage,showpacs,superscriptaddress,floatfix]{revtex4-1}
\usepackage{graphicx,graphics}
\usepackage{dcolumn}
\usepackage{amsmath,amssymb,amsfonts}
\usepackage{latexsym,verbatim}
\usepackage{bm}
\usepackage{bbold}
\usepackage{color}
\usepackage{ulem}
\usepackage[breaklinks=false,colorlinks,citecolor=blue,linkcolor=blue,urlcolor=blue]{hyperref}
\usepackage[abs]{overpic}
\DeclareMathOperator{\H0}{{\mathbf{H}}_0}
\DeclareMathOperator{\j0}{{\it J}_0}

\DeclareMathOperator{\y0}{{\it Y}_0}
\DeclareMathOperator{\h012}{{\it H}_0^{(1-2)}}
\begin{document}
\title{Lippmann-Schwinger theory for two-dimensional plasmon scattering}
\author{Iacopo Torre}
\email{Iacopo.Torre@sns.it}
\affiliation{NEST, Scuola Normale Superiore, I-56126 Pisa,~Italy}
\affiliation{Istituto Italiano di Tecnologia, Graphene Labs, Via Morego 30, I-16163 Genova,~Italy}
\author{Mikhail I. Katsnelson}
\affiliation{Radboud University, Institute for Molecules and Materials, NL-6525 AJ Nijmegen,~The Netherlands}
\author{Alberto Diaspro}
\affiliation{Optical Nanoscopy, Nanophysics, Istituto Italiano di Tecnologia, Via Morego 30, I-16163 Genova,~Italy}
\affiliation{Dipartimento di Fisica, Universit\`a di Genova, Via Dodecaneso 33, I-16146 Genova,~Italy}
\author{Vittorio Pellegrini}
\affiliation{Istituto Italiano di Tecnologia, Graphene Labs, Via Morego 30, I-16163 Genova,~Italy}
\author{Marco Polini}
\affiliation{Istituto Italiano di Tecnologia, Graphene Labs, Via Morego 30, I-16163 Genova,~Italy}
\begin{abstract}
Long-lived and ultra-confined plasmons in two-dimensional (2D) electron systems may provide a sub-wavelength diagnostic tool to investigate localized dielectric, electromagnetic, and pseudo-electromagnetic perturbations. In this Article, we present a general theoretical framework to study the scattering of 2D plasmons against such perturbations in the non-retarded limit. We discuss both parabolic-band and massless Dirac fermion 2D electron systems. Our theory starts from a Lippmann-Schwinger equation for the screened potential in an inhomogeneous 2D electron system and utilizes as inputs analytical long-wavelength expressions for the density-density response function, going beyond the local approximation. We present illustrative results for the scattering of 2D plasmons against a point-like charged impurity and a one-dimensional electrostatic barrier due to a line of charges. Exact numerical results obtained from the solution of the Lippmann-Schwinger equation are compared with approximate results based on the Born and eikonal approximations. The importance of nonlocal effects is finally emphasized. 
\end{abstract}
\maketitle
\section{Introduction}
\label{sect:Intro}
Plasmons, ubiquitous collective charge density oscillations that occur in metals and semiconductors, have been studied for a long time~\cite{Pines_and_Nozieres,Platzman_and_Wolff,Giuliani_and_Vignale}. In particular, plasmons in ultra-clean two-dimensional (2D) electron systems are particularly interesting since they suffer little losses and can be tuned by the electric field effect. At temperatures well below the optical phonon energy scale, plasmons in high-quality GaAs/AlGaAs heterostructures~\cite{pfeiffer_physicaE_2003}, for example, appear as very sharp peaks in inelastic light scattering spectra, displaying intriguing correlation and nonlocal effects at ultra-low electron densities~\cite{hirjibehedin_prb_2002}.

The field of ``2D plasmonics'' has been recently greatly revitalized by real-space investigations of plasmons in supported graphene sheets by means of scanning near-field optical microscopes~\cite{fei_nature_2012,chen_nature_2012}. Experimental investigations of graphene plasmons have also been carried out in graphene sheets encapsulated between hexagonal boron nitride (hBN) crystals~\cite{mayorov_nanolett_2011,mayorov_nanolett_2012,wang_science_2013,taychatanapat_naturephys_2013,kretinin_nanolett_2014,bandurin_science_2016}. These samples display nearly ideal transport characteristics~\cite{mayorov_nanolett_2011,mayorov_nanolett_2012,wang_science_2013,taychatanapat_naturephys_2013,kretinin_nanolett_2014,bandurin_science_2016}, whereby only one scattering mechanisms (i.e.~electron-acoustic phonon scattering~\cite{hwang_prb_2008}) fully determines dc transport times at room temperature, at least for sufficiently large carrier densities. Room-temperature plasmons in hBN-encapsulated graphene sheets have been demonstrated~\cite{woessner_naturemater_2015} to display record-high confinement factors ($\sim 10^{7}$ volume confinement) and lifetimes approaching $1~{\rm ps}$, the latter being solely limited by the weak scattering of electrons against graphene phonons.

In this Article, we are interested in the scattering properties of 2D plasmons in parabolic-band electron gases and encapsulated graphene sheets. To this end, we lay down a Lippmann-Schwinger theory that enables us to calculate complex reflection and transmission coefficients for 2D plasmons impinging on a great variety of localized perturbations. 

Scattering theories for surface plasmon polaritons in noble metals have been introduced in the past~\cite{Shchegrov_prl_1997,Sanchez_prb_1999,Nikitin_prb_2007,Brucoli_prb_2011}. More recently, scattering of graphene plasmons against one-dimensional (1D) defects has been studied in Refs.~\onlinecite{MartinMorenoACSnano2013,Basov2013}. In particular, the impact of electronic quasi-bound states on the scattering properties of plasmons has been recently studied in Ref.~\onlinecite{Basov2016}. Scattering of plasmons in more exotic electron systems has also been considered, for example in Ref.~\onlinecite{Efimkin2012}.

The main difference between these earlier works and the theory presented in this Article is that we use an {\it electrostatic approximation}, instead of solving Maxwell equations. This offers several advantages with respect to previous works: (i) our theory is essentially semi-analytical, requires little numerical effort, and, most importantly, takes into account {\it nonlocal} effects; (ii) we calculate the density-density response function from the knowledge of a {\it microscopic} Hamiltonian, instead of assuming phenomenological models for the spatial dependence of the conductivity profile (as done in {\it all} papers, with the exception of Ref.~\onlinecite{Basov2016}); (iii) we treat on equal footing many different perturbations (not only electrostatic perturbations coupling to the electron density operator); and (iv) we provide a recipe to include exchange and correlation effects {\it beyond} the celebrated random phase approximation (RPA)~\cite{Giuliani_and_Vignale}, as we explain below in Appendix~\ref{app:manybody}. 

The only disadvantage of our approach is that we are unable to describe scattering of plasmons into far-field modes of the electromagnetic field. In 2D electron gases in GaAs/AlGaAs heterostructures and graphene sheets, these dissipative processes are usually very weak as it has been demostrated both theoretically~\cite{MartinMorenoACSnano2013} and experimentally (see, for example, Ref.~\onlinecite{woessner}) for sufficiently confined 2D plasmons. The reason is that the plasmon momenta at play in these electron systems are much larger that the photon momentum $\omega/c$. This implies that coupling to far-field modes of the electromagnetic field can occur only in the presence of extremely sharp defects/perturbations. However, experimentally realized (electrostatic) defects/perturbations for plasmons are smooth and are therefore unable to couple plasmons to photons. The situation is particularly ``extreme'' in hybrid heterostructures containing graphene, hBN, and nearby metal gates~\cite{woessner,gonzalez_naturenano_2016}. A metal gate in close proximity to graphene suppresses the long-range tail of the inter-electron Coulomb interaction, morphing the usual 2D unscreened plasmon~\cite{Giuliani_and_Vignale} with $\omega_{\rm pl}(q) \propto \sqrt{q}$ into an acoustic plasmon mode~\cite{woessner,gonzalez_naturenano_2016} with a phase velocity that is extremely close to the electron Fermi velocity $v_{\rm F}$. For a given illumination frequency $\omega$, such acoustic plasmons have therefore momenta that are much larger that those of unscreened plasmons. Recent experiments~\cite{woessner} where plasmons in such stacks were launched against smooth 1D electrostatic barriers show that our approximation is fully justified and that our theory explains in a fully quantitative fashion experimental data with {\it no} fitting parameters. 

Our manuscript is organized as following. In Sect.~\ref{sect:plasmons} we present a brief overview of how to approach the non-trivial problem of plasmons in inhomogeneous media~\cite{ishmukhametov_pssb_1971,ishmukhametov_FMM_1975} and we introduce two fundamental quantities: a) the proper density-density response function $\tilde{\chi}_{nn}(\bm q,\bm q',\omega)$ and b) the screened potential $V_{\rm sc}(\bm q,\omega)$. In Sect.~\ref{sect:scatteringproblem} we introduce two scattering geometries of interest in this work, which are schematically reported in Fig.~\ref{fig:geometry}, and a Lippmann-Schwinger equation for the screened potential, which automatically fulfils appropriate asymptotic conditions. In Sect.~\ref{sect:tmatrix} we introduce the key quantity of our 2D Lippmann-Schwinger plasmon scattering theory: the transition function $T(\bm q,\theta,\omega)$. 
The latter fully controls the scattering amplitude $f(\theta_{\bm r},\theta,\omega)$ in the geometry in Fig.~\ref{fig:geometry}(a) and reflection $r_{\theta,\omega}$ and transmission $t_{\theta,\omega}$ coefficients in the geometry in Fig.~\ref{fig:geometry}(b). In Sect.~\ref{sect:opticalth} we derive a useful relation between the amplitude of forward scattering and the total scattering cross section, which is known, in the context of electromagnetic scattering, as optical theorem. Paralleling single-particle quantum-mechanical scattering theory~\cite{Sakurai}, in Sections~\ref{sect:bornapprox}, \ref{sect:eikonal}, and~\ref{sect:partialwaves} we present three approximations for the evaluation of the scattering observables: the {\it Born} approximation, the {\it eikonal} approximation, and the {\it method of partial waves}, respectively. Finally, two concrete problems, i.e.~scattering of a plasmon in a 2D parabolic-band electron system against an electrostatic potential generated by i) a point-like charged impurity and ii) a 1D line of charges, are explicitly solved in Sect.~\ref{sect:example}. These are used to compare exact numerical results---obtained from the full solution of the Lippmann-Schwinger equation---with approximate results based on the Born and eikonal approximations. In the second geometry, we also explicitly quantify the impact of nonlocal effects. A summary of our main results and a brief set of conclusions and perspectives is finally reported in Sect.~\ref{sect:conclusions}. The evaluation of the transition function requires exact expressions for the proper density-density response function of an inhomogeneous 2D electron system, which are carefully derived in Appendix~\ref{sect:chi}. We here stress the importance of the results contained in Eqs.~(\ref{eq:M1}),~(\ref{eq:M2}), and~(\ref{eq:M3}): these give the high-frequency behavior of the density-density response function of an {\it inhomogeneous} 2D electron liquid subject to a very general perturbation, up to next-to-leading order in the frequency. In Appendix~\ref{sect:graphene} we extend these results to the case of an inhomogeneous electron liquid hosted in a graphene sheet. Four additional Appendices report a wealth of useful technical details.

\section{Plasmons in inhomogeneous media}
\label{sect:plasmons}
The linear density response $n_{1}(\bm q,\omega)$ induced by an external scalar potential in an electron liquid can be expressed in terms of the screened potential $V_{\rm sc}(\bm q,\omega)$ and the {\it proper} density-density response function~\cite{Giuliani_and_Vignale} according to
\begin{equation}\label{eq:induceddensity}
n_{1}(\bm q,\omega)=\sum_{\bm q'} \tilde{\chi}_{nn}(\bm q,\bm q',\omega)V_{\rm sc}(\bm q',\omega)~.
\end{equation}
The screened potential is in turn related to the induced density via
\begin{equation}\label{eq:densitytopotential}
V_{\rm sc}(\bm q,\omega)=V_{\rm ext}(\bm q,\omega)+\sum_{\bm q'}v(\bm q,\bm q',\omega) n_{1}(\bm q', \omega)~,
\end{equation}
where $V_{\rm ext}(\bm q,\omega)$ is the external potential and $v(\bm q,\bm q',\omega)$ is the Fourier transform of the electron-electron interaction potential. 
For example, $v(\bm q,\bm q',\omega)=\delta_{\bm q, \bm q'}2\pi e^2/[q\bar{\epsilon}(\omega)]$ for a 2D electron system ($q=|\bm q|$) surrounded by a homogeneous and isotropic dielectric, with a frequency-dependent permittivity $\bar{\epsilon}(\omega)$.
The interaction potential $v(\bm q,\bm q',\omega)$ accounts for all screening effects stemming from nearby dielectrics.
Note that we are neglecting retardation effects ($c=\infty$).

Plasmons are self-sustained charge density oscillations that occur in absence of an external field. They correspond to non-trivial solutions of the integral equation
\begin{equation}\label{eq:plasmons}
\sum_{\bm q'}\epsilon(\bm q, \bm q',\omega)V_{\rm sc}(\bm q',\omega)=0~,
\end{equation}
where we have introduced the dynamical dielectric function:
\begin{equation}\label{eq:epsilon}
\epsilon(\bm q, \bm q',\omega)=\delta_{\bm q, \bm q'}-\sum_{\bm q''}v(\bm q,\bm q'',\omega) \tilde{\chi}_{nn}(\bm q'',\bm q',\omega)~.
\end{equation}

For electron systems that are invariant under spatial translations, $\tilde{\chi}_{nn}(\bm q,\bm q',\omega)=\delta_{\bm q, \bm q'}\tilde{\chi}_{nn}(\bm q,\omega)$, $v(\bm q,\bm q',\omega)=\delta_{\bm q, \bm q'}v(\bm q,\omega)$, and Eq.~(\ref{eq:plasmons}) reduces to the familiar equation~\cite{Giuliani_and_Vignale}
\begin{equation}\label{eq:homogeneousplasmons}
\left[1-v(\bm q, \omega)\tilde{\chi}_{nn}(\bm q,\omega)\right]V_{\rm sc}(\bm q,\omega)\equiv \epsilon(\bm q,\omega)V_{\rm sc}(\bm q,\omega)=0~.
\end{equation}
The solutions of Eq.~(\ref{eq:homogeneousplasmons}) for $V_{\rm sc}(\bm q,\omega)$ are delta functions peaked at the zeroes of $\epsilon(\bm q,\omega)$ and correspond to {\it plane waves} in real space.

\section{Lippmann-Schwinger theory for 2D plasmons}
\label{sect:scatteringproblem}
\begin{figure}
\begin{overpic}[width=\linewidth]{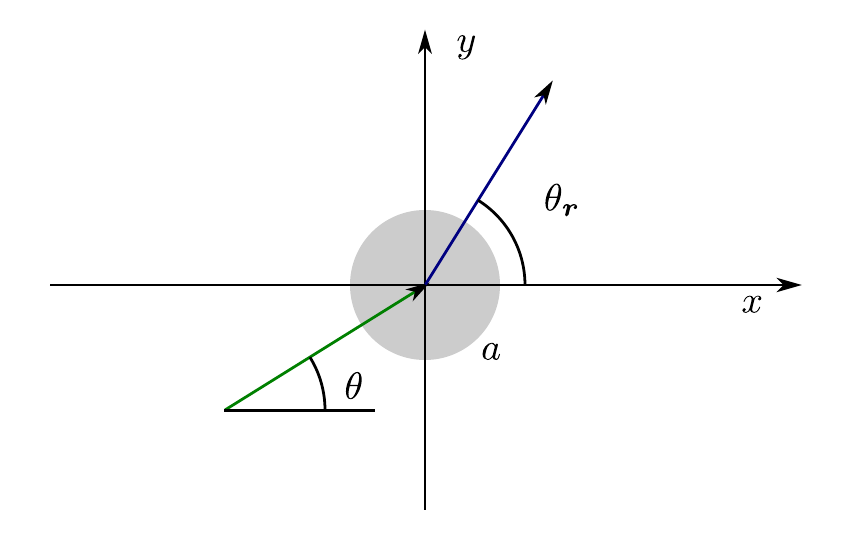}\put(2,150){(a)}\end{overpic}
\begin{overpic}[width=\linewidth]{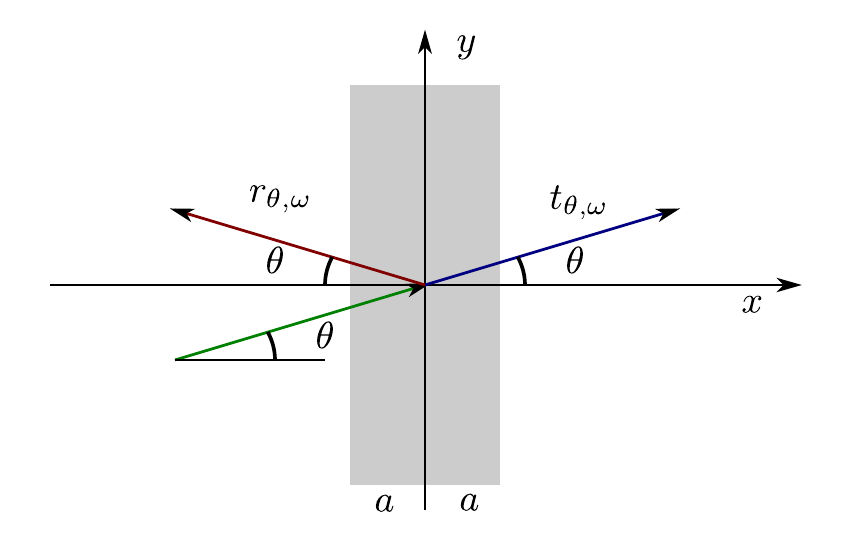}\put(2,150){(b)}\end{overpic}
\caption{\label{fig:geometry} (Color online) The two scattering geometries considered in this work. In panel (a), the perturbation is localized within a circle of radius $a$ (grey-shaded area). In panel (b), the perturbation is confined in the rectangular strip $-a < x < a$ (grey-shaded area). Translational invariance is assumed in the $\hat{\bm y}$ direction. The perturbations $\delta\tilde{\chi}(\bm r, \bm r',\omega)$ and $\delta v(\bm r,\bm r',\omega)$ introduced in the main text in Eqs.~(\ref{eq:chisplittingreal})-(\ref{eq:potentialsplittingreal}) are negligible if either $\bm r$ or $\bm r'$ lie outside the scattering region.}
\end{figure}

We now specialize Eq.~(\ref{eq:plasmons}) to describe the scattering of a plasmon off a spatially-localized inhomogeneity in the 2D electron system. We consider two types of inhomogeneities: (a) one that is localized inside a circle of radius $a$ around the origin and (b) one that is invariant under spatial translations in one direction (the $\hat{\bm y}$ direction) and is confined to a strip of finite width $2a$ in the $\hat{\bm x}$ direction, i.e.~for $-a<x<a$. These are sketched in Fig.~\ref{fig:geometry}(a) and (b), respectively.

We now write the proper response function $\tilde{\chi}_{nn}(\bm q,\bm q',\omega)$ as the sum of a homogeneous part $\tilde{\chi}_{\rm h}(q,\omega)\delta_{\bm q, \bm q'}$ plus a perturbation $\delta \tilde{\chi}(\bm q,\bm q',\omega)$, the latter describing the inhomogeneity present in the 2D electron system:
\begin{equation}\label{eq:chisplitting}
\tilde{\chi}_{nn}(\bm q,\bm q',\omega)=\tilde{\chi}_{\rm h}(q,\omega)\delta_{\bm q, \bm q'}+\frac{1}{S}\delta\tilde{\chi}(\bm q, \bm q',\omega)~.
\end{equation}
Here, $S = L_{x}L_{y}$ is the 2D electron system area. In writing the above equation we assumed, for the sake of simplicity, that the 2D electron system in the absence of perturbations is homogeneous and isotropic---this implies that the homogeneous part $\tilde{\chi}_{\rm h}(q,\omega)$ of the density-density response function depends only on $q = |\bm q|$.

In what follows, we introduce, without loss of generality, the following parametrization of the uniform part of the proper density-density response function:
\begin{equation}\label{eq:chih}
\tilde{\chi}_{\rm h}(q,\omega)=\frac{D}{e^2 \pi} \frac{q^2}{\omega^2} {\cal G}(q,\omega)~,
\end{equation}
where $D$ is the so-called Drude weight~\cite{abedinpour_prb_2011} and ${\cal G}( q,\omega)$ is a correction factor that takes into account all the effects beyond simple Drude theory, including {\it nonlocal} effects. 

Similarly, we split the interaction potential into two parts:
\begin{equation}\label{eq:potentialsplitting}
v(\bm q,\bm q',\omega)=\delta_{\bm q, \bm q'}v(q,\omega)+\frac{1}{S}\delta v(\bm q,\bm q',\omega)~,
\end{equation}
where
\begin{equation}\label{eq:vh}
v(q,\omega)=\frac{2 \pi e^2}{\bar{ \epsilon}(\omega) q}\mathcal{F}(q,\omega)
\end{equation}
represents the homogeneous part of the interaction, which does not depend on the direction of $\bm q$, while $\delta v(\bm q,\bm q',\omega)$ stems from an inhomogeneity in the dielectric environment surrounding the 2D electron system.

In Eq.~(\ref{eq:vh}), $\bar{\epsilon}(\omega)$ is a suitable frequency-dependent permittivity 
and ${\cal F}(q,\omega)$ is a form factor that takes into account deviations from the pure 2D Coulomb law~\cite{Giuliani_and_Vignale} $2\pi e^2/q$. 
These may occur in quantum wells of GaAs/AlGaAs where ${\cal F}$ takes into account the finite thickness of the quantum well and its geometric form~\cite{ando_rmp_1982} or in graphene sheets encapsulated between 
slabs of hBN crystals, where ${\cal F}$ captures effects stemming from the finite 
thickness of hBN~\cite{tomadin_prl_2015}.

In real space, Eqs.~(\ref{eq:chisplitting}) and~(\ref{eq:potentialsplitting}) read as following: 
\begin{equation}\label{eq:chisplittingreal}
\tilde{\chi}_{nn}(\bm r,\bm r',\omega)=\tilde{\chi}_{\rm h}(|\bm r -\bm r'|,\omega)+\frac{1}{S}\delta\tilde{\chi}(\bm r, \bm r',\omega)
\end{equation}
and 
\begin{equation}\label{eq:potentialsplittingreal}
v(\bm r,\bm r',\omega)=v(|\bm r-\bm r'|,\omega)+\frac{1}{S}\delta v(\bm r,\bm r',\omega)~.
\end{equation}
We assume that the perturbations $\delta\tilde{\chi}(\bm r, \bm r',\omega)$ and $\delta v(\bm r,\bm r',\omega)$ are negligible if either $\bm r$ or $\bm r'$ lie outside the scattering region, see Fig.~\ref{fig:geometry}.

We now introduce the homogeneous part of the dielectric function
\begin{eqnarray}\label{eq:epshomdefinition}
\epsilon_{\rm h}(q,\omega) &\equiv & 1-v( q,\omega)\tilde{\chi}_{\rm h}(q,\omega) \nonumber\\
&=& 1-\frac{2D}{\bar{\epsilon}(\omega)\omega^2}q\mathcal{F}(q,\omega)\mathcal{G}(q,\omega)~,
\end{eqnarray}
the inverse of the effective interaction
\begin{eqnarray}\label{eq:Whomdefinition}
W_{\rm h}^{-1}(q,\omega) &\equiv & \frac{\epsilon_{\rm h}(q,\omega)}{v(q,\omega)} = \frac{q\bar{\epsilon}(\omega)}{2\pi e^2 \mathcal{F}(q, \omega)}\nonumber\\
&\times&\left[1-\frac{2D}{\bar{\epsilon}(\omega)\omega^2}q\mathcal{F}(q,\omega)\mathcal{G}(q,\omega)\right]~,
\end{eqnarray}
and the {\it scattering kernel}
\begin{eqnarray}\label{eq:Udefinition}
\Delta(\bm q,\bm q',\omega) &\equiv &
\delta\tilde{\chi}(\bm q, \bm q',\omega)
+\frac{\delta v(\bm q, \bm q',\omega)}{v(q, \omega)}\tilde{\chi}_{\rm h}(q',\omega)\nonumber\\
&+& \frac{1}{S}\sum_{\bm q''}\frac{\delta v(\bm q, \bm q'',\omega)}{v( q, \omega)}\delta \tilde{\chi}(\bm q'',\bm q',\omega)~.
\end{eqnarray}
Note that Eq.~(\ref{eq:Udefinition}) establishes a crucial relationship between scattering theory and microscopic many-body theory, which can be used to calculate the response function $\delta\tilde{\chi}(\bm q, \bm q',\omega)$ of the inhomogeneous 2D electron system that appears in Eq.~(\ref{eq:Udefinition}).

With these definitions, we can rewrite Eq.~(\ref{eq:plasmons}) in the following appealing manner:
\begin{equation}\label{eq:scatteringequation}
W_{\rm h}^{-1}(q,\omega)V_{\rm sc}(\bm q,\omega)=\frac{1}{S}\sum_{\bm q'}\Delta(\bm q,\bm q', \omega)V_{\rm sc}(\bm q',\omega)~.
\end{equation}
Eq.~(\ref{eq:scatteringequation}) closely resembles the momentum-space version of Schr\"odinger's equation for an electron of mass $m$ in a generic {\it nonlocal} potential $U(\bm q,\bm q')$:
\begin{equation}\label{eq:schroedingerscattering}
\left(E-\frac{\hbar^2q^2}{2m}\right)\psi(\bm q)=\frac{1}{S}\sum_{\bm q'}U(\bm q,\bm q')\psi(\bm q')~.
\end{equation}
Comparing Eq.~(\ref{eq:scatteringequation}) with Eq.~(\ref{eq:schroedingerscattering}), we clearly see that $W_{\rm h}^{-1}(q,\omega)$ plays the role of $E-\hbar^2 q^2/(2m)$, $\Delta(\bm q, \bm q',\omega)$ plays the role of the scattering potential $U(\bm q,\bm q')$, and $V_{\rm sc}(\bm q,\omega)$ is the analogue of the wavefunction $\psi(\bm q)$.

In the following, we assume that the unperturbed system has a single plasmon mode at a given frequency $\omega$. This means that $\epsilon_{\rm h}(q,\omega)$ has only one zero as a function of $q$ for any given $\omega$. This allows us to unambiguously define the {\it plasmon wavevector} of the homogeneous system 
as the solution $q_{\rm pl}=q_{\rm pl}(\omega)$ of the following equation:
\begin{equation}\label{eq:kpdefinition}
\epsilon_{\rm h}(q_{\rm pl}, \omega)=0~.
\end{equation}
Our theory can be easily extended to anisotropic media, for which $q_{\rm pl}$ depends on the propagation direction, and to take into account the presence of multiple plasmon modes at a given frequency. 

For the geometry in Fig.~\ref{fig:geometry}(a) and just as in the case of single-particle quantum-mechanical scattering theory~\cite{Sakurai}, we are interested in solutions of Eq.~(\ref{eq:scatteringequation}) whose {\it asymptotic} behavior is given by the sum of an incoming plane wave plus a scattered wave:
\begin{equation}\label{eq:asymptoticbehaviour}
V_{\rm sc}(\bm r,\omega) \simeq e^{i \bm q_{\rm pl} \cdot \bm r} + \frac{e^{i q_{\rm pl} r}}{\sqrt{r}}f(\theta_{\bm r},\theta,\omega)~,
\end{equation}
where $\bm q_{\rm pl}\equiv q_{\rm pl}[\hat{\bm x}\cos(\theta)+\hat{\bm y}\sin(\theta)]$, $\theta$ is the polar angle of the wavevector of the incoming wave, $\theta_{\bm r}$ is the polar angle of $\bm r$, and $f(\theta_{\bm r},\theta,\omega)$ is the scattering amplitude induced by the inhomogeneity.

For the geometry in Fig.~\ref{fig:geometry}(b), which is translationally invariant along the $\hat{\bm y}$ direction, the required asymptotic behavior is given by
\begin{equation}\label{eq:asymptoticbehaviour1d}
\begin{split}
V_{\rm sc} & (\bm r, \omega ) \simeq e^{i  q_{\rm pl} \sin(\theta) y}\\
& \times
\begin{cases}
e^{i  q_{\rm pl} \cos(\theta) x}+r_{\theta, \omega}e^{-i  q_{\rm pl} \cos(\theta) x},~{\rm for}~x \to -\infty\\
t_{\theta, \omega}e^{i q_{\rm pl} \cos (\theta) x},~{\rm for}~x \to +\infty
\end{cases}
\end{split}~,
\end{equation}
where $\theta$ is the angle between the wavevector of the incoming wave and the $\hat{\bm x}$ axis, while $r_{\theta, \omega}$ and $t_{\theta, \omega}$ are the reflection and transmission coefficients, respectively.

The asymptotic behaviors (\ref{eq:asymptoticbehaviour}) and~(\ref{eq:asymptoticbehaviour1d}) can be more easily enforced using a formalism {\it \`a la} Lippmann-Schwinger~\cite{Sakurai}. Indeed, we claim that a solution of 
\begin{eqnarray}\label{eq:lippmannschwinger}
V_{\rm sc}(\bm q,\omega) &=&V^{(0)}(\bm q,\omega) + W_{\rm h}(q,\omega)\nonumber\\
&\times& \frac{1}{S} \sum_{\bm q'} \Delta(\bm q, \bm q',\omega)V_{\rm sc}(\bm q',\omega)~,
\end{eqnarray}
with $V^{(0)}(\bm q, \omega)$ satisfying
\begin{equation}\label{eq:incomingwave}
\epsilon_{\rm h}( q,\omega)V^{(0)}(\bm q,\omega) = 0
\end{equation}
and $W_{\rm h}(q,\omega)$ satisfying the {\it distributional} equation
\begin{equation}\label{eq:effectiveinteraction}
W_{\rm h}(q,\omega)\frac{\epsilon_{\rm h}(q,\omega)}{v(q,\omega)}=1~,
\end{equation}
is also a solution of Eq.~(\ref{eq:scatteringequation}). 

To prove our assertion, it is sufficient to multiply Eq.~(\ref{eq:lippmannschwinger}) by $W_{\rm h}^{-1}(q,\omega)$ and use Eqs.~(\ref{eq:incomingwave})-(\ref{eq:effectiveinteraction}). Eq.~(\ref{eq:lippmannschwinger}) is the desired Lippmann-Schwinger equation for 2D plasmon scattering.

The solutions of Eqs.~(\ref{eq:incomingwave})-(\ref{eq:effectiveinteraction}) are not unique.
To impose the asymptotic conditions (\ref{eq:asymptoticbehaviour})-(\ref{eq:asymptoticbehaviour1d}), we choose: 
i) $V^{(0)}$ as a delta function in wavevector space (a plane wave in real space):
\begin{equation}\label{eq:v0}
V^{(0)}(\bm q,\omega)=(2 \pi)^2 \delta(\bm q-\bm q_{\rm pl})~,
\end{equation}
which corresponds to the first term in Eqs.~(\ref{eq:asymptoticbehaviour})-(\ref{eq:asymptoticbehaviour1d}), and ii) the solution of Eq.~(\ref{eq:effectiveinteraction}) corresponding to an outgoing cylindrical wave. 
As shown in Appendix~\ref{app:effectiveinteraction}, the solution of Eq.~(\ref{eq:effectiveinteraction}) corresponding to an outgoing  cylindrical wave is:
\begin{eqnarray}\label{eq:whom+}
W_{\rm h}^{(+)}(q,\omega)&\equiv &\frac{1}{W_{\rm h}^{-1}(q,\omega)+i0^+} \nonumber\\
&=&\mathcal{P}\frac{1}{W_{\rm h}^{-1}(q,\omega)}- i \pi C(\omega) \delta(q-q_{\rm pl})~,
\end{eqnarray}
where
\begin{eqnarray}\label{eq:resdefinition}
C(\omega) &=& \lim_{q \to q_{\rm pl}}\frac{q_{\rm pl}-q}{W_{\rm h}^{-1}(q,\omega)} \nonumber\\
&=&\frac{2 \pi e^2 \mathcal{F}(q_{\rm pl},\omega)}{\bar{\epsilon}(\omega)\left[1+q_{\rm pl}\frac{\mathcal{F}'(q_{\rm pl},\omega)}{\mathcal{F}(q_{\rm pl},\omega)}+q_{\rm pl}\frac{\mathcal{G}'(q_{\rm pl},\omega)}{\mathcal{G}(q_{\rm pl},\omega)}\right]}~.
\end{eqnarray}
Here, ${\cal F}'(q,\omega) \equiv \partial {\cal F}(q,\omega)/\partial q$ and ${\cal G}'(q,\omega) \equiv \partial {\cal G}(q,\omega)/\partial q$.

With these definitions we can separate the effective interaction into a {\it universal} function of $q$ plus a correction ${\cal W}(q,\omega)$:
\begin{equation}
\begin{split}
&W_{\rm h}^{(+)}(q,\omega)= \\
& =\frac{C(\omega)}{q_{\rm pl}}\left[\mathcal{P}\frac{1}{1-\frac{q}{q_{\rm pl}}}-i\pi \delta\left(1-\frac{q}{q_{\rm pl}}\right)+\frac{q_{\rm pl}}{q}+\mathcal{W}(q,\omega)\right]~,
\end{split}
\end{equation}
where
\begin{widetext}
\begin{equation}\label{eq:Wcorrection}
\mathcal{W}(q,\omega)  
=\frac{\mathcal{F}(q,\omega)\left[\mathcal{F}(q_{\rm pl},\omega)\mathcal{G}(q_{\rm pl},\omega)+q_{\rm pl}\mathcal{F}'(q_{\rm pl},\omega)\mathcal{G}(q_{\rm pl},\omega)+q_{\rm pl}\mathcal{F}(q_{\rm pl},\omega)\mathcal{G}'(q_{\rm pl},\omega)\right]}{\mathcal{F}(q_{\rm pl},\omega)\frac{q}{q_{\rm pl}}\left[\mathcal{F}(q_{\rm pl},\omega)\mathcal{G}(q_{\rm pl},\omega)-\frac{q}{q_{\rm pl}}\mathcal{F}(q,\omega)\mathcal{G}(q,\omega)\right]}-\frac{q_{\rm pl}}{q_{\rm pl}-q}-\frac{q_{\rm pl}}{q}~.
\end{equation}
\end{widetext}
Note that ${\cal W}(q,\omega)$ vanishes identically if {\it both} ${\cal F}(q,\omega)$ and ${\cal G}(q,\omega)$ are set to one.

Since the intensity associated with a travelling plasmon with a fixed wavevector is proportional to the square of its potential, the information about the flux of energy is carried by the modulus square of the scattering amplitude $|f(\theta_{\bm r},\theta,\omega)|^2$ in the geometry of Fig~\ref{fig:geometry}(a) and by the square modulus of $r_{\theta,\omega}$ and $t_{\theta,\omega}$ in the geometry of Fig~\ref{fig:geometry}(b).
More precisely the ratio between the amount of power scattered into a small angle $d\theta$ around $\theta_{\bm r}$ and the intensity (i.e.~power per unit length) of the incoming wave is $|f(\theta_{\bm r},\theta,\omega)|^2d\theta$. The total scattered power divided by the intensity of the original wave is given by the total cross section
\begin{equation}\label{eq:crossectiondefinition}
\Sigma(\theta,\omega)\equiv \int_{-\pi}^\pi d\theta_{\bm r} |f(\theta_{\bm r},\theta,\omega)|^2~.
\end{equation}
In rotationally invariant system $\Sigma(\theta,\omega)$ obviously does not depend on the angle $\theta$ and is a function of $\omega$ only.

\section{Transition Function}
\label{sect:tmatrix}
We now turn to relate the scattering amplitude $f(\theta_{\bm r},\theta,\omega)$ and the reflection and transmission coefficients $r_{\theta, \omega}$ and $t_{\theta, \omega}$ to the solutions of Eq.~(\ref{eq:lippmannschwinger}). To this end, it is useful to define the {\it transition function} $T$. 
Since the final results are slightly different for the two scattering geometries in Figs.~\ref{fig:geometry}(a) and (b), we split the discussion into two separate parts, in Sects.~\ref{subsect:geometry-a} and~\ref{subsect:geometry-b}, respectively.
\subsection{Geometry in Fig.~\ref{fig:geometry}(a)} 
\label{subsect:geometry-a}
In this geometry, the transition function is defined by
\begin{equation}\label{eq:tmatrixdefinition}
T(\bm q,\theta,\omega) \equiv \frac{1}{S}\sum_{\bm q'}\Delta(\bm q, \bm q',\omega)V_{\rm sc}(\bm q',\omega)~,
\end{equation}
where $V_{\rm sc}(\bm q,\omega)$ is the solution of Eq.~(\ref{eq:lippmannschwinger}) with $V^{(0)}(\bm q,\omega)$ given by (\ref{eq:v0}) and effective interaction given by (\ref{eq:whom+}).
Note that $T$ is a function of a reciprocal vector $\bm q$, of an angle $\theta$ giving the direction of the incoming wave, and of the frequency $\omega$.

The transition function satisfies the following integral equation
\begin{equation}\label{eq:tmatrixequation}
\begin{split}
T(\bm q,\theta,\omega) & =  \Delta(\bm q,\bm q_{\rm pl},\omega)\\
& +\frac{1}{S}\sum_{\bm q'} \Delta(\bm q, \bm q',\omega)W_{\rm h}^{(+)}(q',\omega)T(\bm q',\theta,\omega)~,
\end{split}
\end{equation}
as one can easily verify by inserting Eq.~(\ref{eq:lippmannschwinger}) in Eq.~(\ref{eq:tmatrixdefinition}). 

To make a link between the transition function and the scattering amplitude $f(\theta_{\bm r},\theta,\omega)$ we must consider the asymptotic behavior of Eq.~(\ref{eq:lippmannschwinger}) in real space. This is carefully considered in Appendix~\ref{app:realspace}. The final result is
\begin{equation}\label{eq:scatteringamplitude}
f(\theta_{\bm r},\theta,\omega) =-\frac{e^{i\frac{\pi}4}}{\sqrt{2\pi}}\sqrt{q_{\rm pl}}C(\omega) T(q_{\rm pl}\hat{\bm r},\theta,\omega)~.
\end{equation}
\subsection{Geometry in Fig.~\ref{fig:geometry}(b)}
\label{subsect:geometry-b}

Since this geometry is translationally invariant in the $\hat{\bm y}$ direction, we can rewrite the scattering kernel (\ref{eq:Udefinition}) as
\begin{equation}\label{eq:U1d}
\Delta(\bm q,\bm q',\omega) = 2\pi\delta(q_y-q_y') \Delta(q_x, q_x',q_y,\omega)~.
\end{equation}
Eq.~(\ref{eq:lippmannschwinger}) then becomes
\begin{eqnarray}\label{eq:lippmannschwinger1d0}
V_{\rm sc} (q_x,q_y,\omega) & = & V^{(0)}(q_x,q_y,\omega)+W_{\rm h}^{(+)}\left(\sqrt{q_x^2+q_y^2},\omega \right)\nonumber\\
& \times& \frac{1}{L_{x}}\sum_{q_x'} \Delta(q_x,q_x',q_y,\omega)V_{\rm sc}(q_x',q_y,\omega)~,\nonumber\\
\end{eqnarray}
where $L_{x}$ has been defined after Eq.~(\ref{eq:chisplitting}).

We now separate the two components of Eq.~(\ref{eq:v0}):
\begin{eqnarray}
V^{(0)}(q_x,q_y) &=& 2\pi \delta(q_y-q_{\rm pl}\sin(\theta)) \nonumber\\
&\times& 2\pi \delta(q_x-q_{\rm pl}\cos(\theta))~.
\end{eqnarray}
Because of translational invariance in the $\hat{\bm y}$ direction, we can take the solution to have the form
\begin{equation}\label{eq:ansatz}
V_{\rm sc}(q_x,q_y,\omega)= 2\pi \delta(q_y-q_{\rm pl}\sin(\theta))
V_{\rm sc}(q_x,\theta,\omega)~.
\end{equation}
The Lippmann-Schwinger equation (\ref{eq:lippmannschwinger}) then becomes
\begin{eqnarray}\label{eq:lippmannschwinger1d}
& V_{\rm sc} (q_x,\theta,\omega)   =  2\pi \delta(q_x-q_{\rm pl}\cos\theta) \nonumber\\
& +  W_{\rm h}^{(+)}\left(\sqrt{q_x^2+q_{\rm pl}^2\sin^2\theta},\omega\right)\nonumber\\
& \times  \frac{1}{L_x}\sum_{q_x'} \Delta(q_x,q_x',q_{\rm pl}\sin\theta,\omega)V_{\rm sc}(q_x',\theta,\omega)~.
\end{eqnarray}

In analogy to Eq.~(\ref{eq:tmatrixdefinition}), we define
\begin{equation}\label{eq:tmatrixdefinition1d}
T(q_x,\theta,\omega) \equiv \frac{1}{L_x}\sum_{q_x'} \Delta(q_x, q_x',q_{\rm pl}\sin(\theta),\omega)V_{\rm sc}(q_x',\theta,\omega)~,
\end{equation}
which satisfies the following integral equation:
\begin{widetext}
\begin{equation}\label{eq:tmatrixequation1d}
\begin{split}
T(q_x,\theta,\omega)  =  \Delta(q_x, q_{\rm pl}\cos(\theta), q_{\rm pl}\sin(\theta),\omega) +\frac{1}{L_x}\sum_{ q_x'} \Delta(q_x, q_x',q_{\rm pl}\sin(\theta),\omega)  W_{\rm h}^{(+)}\left(\sqrt{q_x'^2+q_{\rm pl}^2\sin^2(\theta)},\omega\right)T(q_x',\theta,\omega)~.
\end{split}
\end{equation}
\end{widetext}
This is the equation we solved numerically in the geometry (b) described in Section~\ref{sect:example}.

Using the asymptotic behavior of Eq.~(\ref{eq:lippmannschwinger1d})---see Appendix~\ref{app:realspace}---and comparing the result to Eq.~(\ref{eq:asymptoticbehaviour1d}), we finally obtain:
\begin{equation}\label{eq:transcoeff}
t_{\theta, \omega} = 1-i\frac{C(\omega)}{\cos(\theta)}T(q_{\rm pl}\cos(\theta),\theta,\omega)
\end{equation}
and
\begin{equation}\label{eq:refcoeff}
r_{\theta, \omega}=-i\frac{C(\omega)}{\cos(\theta)}T(-q_{\rm pl}\cos(\theta),\theta,\omega)~.
\end{equation}
In the theory of single-particle quantum-mechanical scattering~\cite{Sakurai} the analytical continuation of the transition function into the upper half of the complex plane can display poles at purely imaginary values of the wavevector  corresponding to the energies of single-particle bound states. Similarly, for the case of plasmon scattering, a localized plasmon resonance---i.e.~a solution of Eq.~(\ref{eq:plasmons}) that decays exponentially, in real space, far from the perturbation---manifests as a pole in the transition function $T(q_x,\theta,\omega)$ at a purely imaginary value of $q_x$. 

\section{Optical theorem}
\label{sect:opticalth}
In this Section we derive a useful relation between the amplitude of forward scattering (i.e.~scattering in the same direction of the incoming wave) and the total scattering cross section. In the context of electromagnetic scattering, this is known as ``optical theorem''. It holds if dissipation can be neglected during the scattering process. Once again, we split the derivation into two parts, depending on the scattering geometry.
\subsection{Geometry in Fig.~\ref{fig:geometry}(a)}
\label{subsect:optical_theorem_geometry_a}
Using Eq.~(\ref{eq:tmatrixdefinition}), we can write the imaginary part of the transition function for the forward scattering process as 
\begin{equation}\label{eq:ImTforward}
\begin{split}
\Im m\left\{T(\bm q_{\rm pl},\theta,\omega)\right\} =\Im m\left\{\frac{1}{S}\sum_{\bm q'}\Delta(\bm q_{\rm pl}, \bm q',\omega)V_{\rm sc}(\bm q',\omega)\right\}\\
=\Im m\left\{\frac{1}{S}\sum_{\bm q'}\int d^2 \bm q\, \delta(\bm q -\bm q_{\rm pl})\Delta(\bm q, \bm q',\omega)V_{\rm sc}(\bm q',\omega)\right\}~.
\end{split}
\end{equation}
We can rewrite the delta function using the complex conjugate of the Lippmann-Schwinger equation (\ref{eq:lippmannschwinger}), and Eqs.~(\ref{eq:v0}), (\ref{eq:whom+}), and~(\ref{eq:tmatrixdefinition}):
\begin{equation}\label{eq:auxiliary-delta-function}
\begin{split}
& \delta(\bm q-\bm q_{\rm pl})=\frac{1}{(2\pi)^2}V_{\rm sc}^*(\bm q,\omega)\\
& -\frac{1}{(2\pi)^2}\left[\mathcal{P}\frac{1}{W_{\rm h}^{-1}(q,\omega)}+ i \pi C(\omega) \delta(q-q_{\rm pl})\right]T^*(\bm q,\theta,\omega)~.
\end{split}
\end{equation}
Substituting Eq.~(\ref{eq:auxiliary-delta-function}) into Eq.~(\ref{eq:ImTforward}) we get
\begin{equation}
\begin{split}
& \Im m\left\{T(\bm q_{\rm pl},\theta,\omega)\right\} =-\frac{ C(\omega)}{4\pi}\int d\bm q|T(\bm q,\theta,\omega)|^2\delta(q-q_{\rm pl})\\
&+\Im m\left\{\int \frac{d\bm q}{(2\pi)^2}\int \frac{d\bm q'}{(2\pi)^2}V_{\rm sc}^*(\bm q,\omega)\Delta(\bm q, \bm q',\omega)V_{\rm sc}(\bm q',\omega)\right\}\\
& =-\frac{C(\omega)q_{\rm pl}}{4\pi}\int d\theta'| T(q_{\rm pl}[\hat{\bm x} \cos \theta'+\hat{\bm y} \sin \theta' ],\theta,\omega)|^2~.
\end{split}
\end{equation}
The term in the second line of the previous equation is proportional to the power absorbed by the inhomogeneous electron system~\cite{Giuliani_and_Vignale} and can therefore be neglected if dissipation is small.

Making use of Eqs.~(\ref{eq:crossectiondefinition}-\ref{eq:scatteringamplitude}) this relation can be recast in the form of an optical theorem~\cite{Sakurai}
\begin{equation}\label{eq:opticaltheorem}
\Im m\left\{\frac{2\sqrt{2\pi}e^{-i\frac{\pi}{4}}}{\sqrt{q_{\rm pl}}}f(\theta,\theta,\omega)\right\}
=\Sigma(\theta,\omega)~.
\end{equation}
\subsection{Geometry in Fig.~\ref{fig:geometry}(b)}
\label{subsect:optical_theorem_geometry_b}
Following the same steps as those in Sect.~\ref{subsect:optical_theorem_geometry_a}, we derive a very similar relation for the forward-scattering transition function in the geometry sketched in Fig.~\ref{fig:geometry}(b):
\begin{equation}
\begin{split}
&\Im m\left\{T(q_{\rm pl} \cos\theta,\theta,\omega)\right\}=\\
&-\frac{C(\omega)}{2}\int dq_x |T(q_x,\theta,\omega)|^2\delta \left(\sqrt{q_{\rm pl}^2 \sin^2\theta-q_x^2}-q_{\rm pl}\right)~.
\end{split}
\end{equation}
Using the definitions in Eqs.~(\ref{eq:transcoeff}) and~(\ref{eq:refcoeff}) we get
\begin{equation}\label{eq:opticlatheorem1d}
|r_{\theta, \omega}|^2+|t_{\theta,\omega}|^2=1~.
\end{equation}
The latter simply expresses conservation of energy in absence of dissipation.

\section{The Born approximation}
\label{sect:bornapprox}

In this Section we discuss the Born approximation for the two geometries of interest in this work.

\subsection{Geometry in Fig.~\ref{fig:geometry}(a)}

Eq.~(\ref{eq:tmatrixequation}) can be handled exactly in a numerical fashion, as we will discuss below, provided that the scattering kernel $\Delta(\bm q,\bm q',\omega)$ is known. In this Section, however, we wish to introduce an approximate perturbative approach in powers of $\Delta(\bm q,\bm q',\omega)$, which is usually termed ``Born approximation'' in ordinary single-particle quantum-mechanical scattering theory~\cite{Sakurai}.

We start by writing the transition function as a power series:
\begin{equation}\label{eq:tmatrixseries}
T(\bm q,\theta,\omega)\equiv \sum_{n=1}^\infty\lambda^n T^{(n)}(\bm q,\theta,\omega)~,
\end{equation}
where $\lambda$ is a dimensionless bookkeeping parameter, which will be set to unity at the end of calculation. We also multiply the kernel $\Delta(\bm q,\bm q',\omega)$ in Eq.~(\ref{eq:tmatrixequation}) by the same parameter $\lambda$.
The equation for the transition function becomes
\begin{equation}\label{eq:tmatrixequationpert}
\begin{split}
& \sum_{n=1}^\infty\lambda^n T^{(n)}(\bm q,\theta,\omega) =\lambda \Delta(\bm q, \bm q_{\rm pl},\omega)\\
& +\frac{1}{S}\sum_{\bm q'} \lambda \Delta(\bm q, \bm q',\omega)W_{\rm h}^{(+)}( q',\omega)\sum_{n=1}^\infty\lambda^n T^{(n)}(\bm q',\theta,\omega)~.
\end{split}
\end{equation}
Collecting terms that appear in Eq.~(\ref{eq:tmatrixequationpert}) with the same power of $\lambda$ and setting $\lambda=1$, we finally obtain
\begin{equation}\label{eq:tmatrixFOBA}
T^{(1)}(\bm q,\theta,\omega) =  \Delta(\bm q,\bm q_{\rm pl},\omega)
\end{equation}
and
\begin{equation}\label{eq:tmatrixBA}
\begin{split}
& T^{(n+1)}(\bm q,\theta,\omega) \\
& =\frac{1}{S}\sum_{\bm q'} \Delta(\bm q, \bm q',\omega)W_{\rm h}^{(+)}(q',\omega)T^{(n)}(\bm q',\theta,\omega)~,
\end{split}
\end{equation}
for $n\geq1$. This series yields a scattering amplitude
\begin{equation}\label{eq:scatteringamplitudeBA}
\begin{split}
f(\theta_{\bm r},\theta,\omega) & =\sum_{n=1}^\infty f^{(n)}(\theta_{\bm r},\theta,\omega)\\
&=- \frac{e^{i\frac{\pi}4}}{\sqrt{2\pi}}\sqrt{q_{\rm pl}} C(\omega)\sum_{n=1}^\infty T^{(n)}(q_{\rm pl}\hat{\bm r},\theta,\omega)~.
\end{split}
\end{equation}
As one can see from Eq.~(\ref{eq:tmatrixFOBA}), the leading term of the expansion in Eq.~(\ref{eq:tmatrixseries}) is particularly easy to calculate since it is simply given by the scattering kernel evaluated at the outcoming and incoming wavevectors:
\begin{equation}\label{eq:FOBA}
f^{(1)}(\theta_{\bm r},\theta,\omega)=- \frac{e^{i\frac{\pi}4}}{\sqrt{2\pi}}\sqrt{q_{\rm pl}} C(\omega) \Delta(q_{\rm pl}\hat{\bm r},\bm q_{\rm pl},\omega)~.
\end{equation}
Eq.~(\ref{eq:FOBA}) represents the first-order Born approximation for the scattering amplitude and often represents a good starting tool to understand, at least qualitatively, the behavior of 2D plasmon scattering in a purely analytical fashion.

We note that the scattering amplitude calculated using the Born approximation does not fulfil the optical theorem~(\ref{eq:opticaltheorem}), order by order. For example, the right-hand side of Eq.~(\ref{eq:opticaltheorem}) calculated with $f(\theta',\theta,\omega)$ at the level of the first-order Born approximation is equal to the left-hand side of Eq.~(\ref{eq:opticaltheorem}) with $f(\theta,\theta,\omega)$ calculated using the second-order Born approximation.

A natural question that arises at this point is when the Born series converges and when it is legitimate to keep only the first terms of the series. The Born approximation works well when the difference between the full solution for the screened potential $V_{\rm sc}(\bm r,\omega)$ and the incoming wave $V^{(0)}(\bm r,\omega)$ within the scattering region is small. Looking at the real-space formulation of the Lippmann-Schwinger equation in Appendix~\ref{app:realspace}, we can write the difference between the full solution and the incoming wave for $\bm r \approx {\bm 0}$ as
\begin{equation}\label{eq:bornvalidity}
\begin{split}
&|V_{\rm sc}(\bm r \approx \bm 0,\omega)-V^{(0)}(\bm r \approx \bm 0,\omega)|\\
& \approx\left|\int d \bm r'W_{\rm h}^{(+)}(|\bm r'|,\omega) \int d \bm r'' \frac{1}{S}\Delta(\bm r',\bm r'',\omega)V_{\rm sc}(\bm r'',\omega)\right|\\
& \approx \left|\int d \bm r'W_{\rm h}^{(+)}(|\bm r'|,\omega) \int d \bm r'' \frac{1}{S}\Delta(\bm r',\bm r'',\omega)V^{(0)}(\bm r'',\omega)\right|\\
& =\left|\frac{1}{S}\sum_{\bm q}W_{\rm h}^{(+)}(q,\omega)\Delta(\bm q,\bm q_{\rm pl},\omega)\right|~.
\end{split}
\end{equation}
If the above quantity is much smaller than unity, the perturbative series converges and the Born approximation is good.
\subsection{Geometry in Fig.~\ref{fig:geometry}(b)}
In the case of the geometry in Fig.~\ref{fig:geometry}(b), we can still express the transition function as a power series
\begin{equation}\label{eq:tmatrixseries1d}
T(q_x,\theta,\omega)=\sum_{n=1}^\infty T^{(n)}(q_x,\theta,\omega)
\end{equation}
with coefficients given by
\begin{equation}\label{eq:tmatrixFOBA1d}
T^{(1)}(q_x,\theta,\omega) =  \Delta(q_x,q_{\rm pl}\cos(\theta),q_{\rm pl}\sin(\theta),\omega)
\end{equation}
and
\begin{equation}\label{eq:tmatrixBA1d}
\begin{split}
& T^{(n+1)}(q_x,\theta,\omega) =\frac{1}{L_x}\sum_{ q_x'} \Delta(q_x, q_x',q_{\rm pl}\sin(\theta),\omega)\\
& 
\times W_{\rm h}^{(+)}\left(\sqrt{q_{\rm pl}^2\sin^2(\theta)+q_x'^2},\omega\right)T^{(n)}( q'_x,\theta,\omega) ~.
\end{split}
\end{equation}
Transmission and reflection coefficients in the first-order Born approximation read as following: 
\begin{equation}\label{eq:tFOBA}
t_{\theta, \omega}^{(1)}=1-i\frac{C(\omega)}{\cos\theta}\Delta(q_{\rm pl}\cos\theta,q_{\rm pl}\cos\theta,q_{\rm pl}\sin\theta,\omega)
\end{equation}
and
\begin{equation}\label{eq:rFOBA}
r_{\theta, \omega}^{(1)}=-i\frac{C(\omega)}{\cos\theta}\Delta(-q_{\rm pl}\cos\theta,q_{\rm pl}\cos\theta,q_{\rm pl}\sin\theta,\omega)~.
\end{equation}
In general, these expressions do not respect the conservation law~(\ref{eq:opticlatheorem1d}), often leading to the unphysical result $|t_{\theta,\omega}^{(1)}|>1$. For this reason, we prefer to extract the amplitude of the transmission coefficient by using Eq.~(\ref{eq:opticlatheorem1d}), with the reflection coefficient being extracted from Eq.~(\ref{eq:rFOBA}).

Following the same steps as in Eq.~(\ref{eq:bornvalidity}), we obtain a similar convergence criterion:
\begin{equation}
\begin{split}
& \Big| \frac{1}{L_x}\sum_{q_x}W_{\rm h}^{(+)}\left(\sqrt{q_x^2+q_{\rm pl}^2 \sin^2(\theta)},\omega\right)\\
&\times \Delta(q_x,q_{\rm pl} \cos(\theta),q_{\rm pl} \sin(\theta),\omega)\Big| \ll 1~.
\end{split}
\end{equation}
\section{The Eikonal Approximation}
\label{sect:eikonal}
In the geometry depicted in Fig.~\ref{fig:geometry}(b) it is possible to introduce the simplest approximation of the full scattering theory, i.e.~the ``Eikonal approximation''. The latter allows the calculation of the phase of the transmission coefficient. This is the most important scattering observable in all situations in which reflection is small (i.e.~when $|r_{\theta,\omega}|\ll 1$).

The eikonal approximation does not rely on the smallness of the scattering kernel $\Delta(\bm q,\bm q',\omega)$ but requires the plasmon wavelength $2\pi/q_{\rm pl}$ to be much smaller than the lengthscale over which the properties of the inhomogeneous electron liquid vary appreciably.

We lay down the derivation of this approximation under the two simplifying assumptions, which can be relaxed if necessary: (i) $\delta v(\bm q,\bm q',\omega)\equiv 0$ and (ii) $\tilde{\chi}_{nn}(\bm q, \bm q',\omega)=\bm q \cdot \bm q' D_{\bm q-\bm q'}/(Se^2\pi\omega^2)$, with $D_{\bm q}= 2\pi \delta(q_y)D_{q_x}$. Physically,  (ii) derives from the assumption of a local conductivity model $\sigma(x) = i D(x)/(\pi \omega)$, with a Drude weight that changes spatially only along the $\hat{\bm x}$ direction---see Fig.~\ref{fig:geometry}(b).

We start from Eq.~(\ref{eq:plasmons}) and use Eq.~(\ref{eq:epsilon}) and assumptions (i) and (ii). 
Dividing by $v(q,\omega)$, we obtain
\begin{equation}
\begin{split}
& \frac{V_{\rm sc}(q_x,\theta,\omega)}{v(q,\omega)}=\\
& =\frac{1}{\pi e^2\omega^2}\int\frac{dq_x'}{2\pi}(q_xq_x'+q_{\rm pl}^2\sin^2 \theta)D_{q_x-q_x'}V_{\rm sc}(q_x',\theta,\omega)~.
\end{split}
\end{equation}
In real space, the previous equation becomes
\begin{equation}\label{eq:eikonalrealspace}
\begin{split}
&\int dx' v^{-1} (x-x',q_{\rm pl}\sin \theta,\omega) V_{\rm sc}(x',\theta,\omega)=\\
&=\frac{1}{\pi e^2\omega^2}\{-\partial_x[D(x)\partial_x V_{\rm sc}(x,\theta,\omega)]\\
&+q_{\rm pl}^2\sin^2 \theta D(x)V_{\rm sc}(x,\theta,\omega)\}  ~,
\end{split}
\end{equation}
where
\begin{equation}\label{eq:inversepotential}
v^{-1}(x,q_y,\omega) \equiv \int \frac{dq_x}{2\pi} \frac{e^{iq_xx}}{v(q,\omega)}
\end{equation}
with $q = \sqrt{q_x^2+q_y^2}$. We now introduce in Eq.~(\ref{eq:eikonalrealspace}) the eikonal ansatz:
\begin{equation}\label{eq:eikonalansatz}
V_{\rm sc}(x,\theta,\omega)=\exp[i q_{\rm pl}\cos \theta S(x)]~.
\end{equation}
Our target is to derive an equation for the quantity $S(x)$. We find
\begin{eqnarray}\label{eq:eikonalequation}
&&\int dx' v^{-1}(x-x',q_{\rm pl}\sin \theta,\omega)e^{-i q_{\rm pl}\cos \theta [ S(x)-S(x')]} =\nonumber\\
&&\frac{1}{\pi e^2\omega^2}\left\{q_{\rm pl}^2\sin^2 \theta D(x)-e^{-i q_{\rm pl}\cos \theta S(x)}\right.\nonumber\\
&\times& \left.\partial_x\left[D(x)\partial_x e^{i q_{\rm pl}\cos \theta S(x)}\right]\right\}~.
\end{eqnarray}
No approximation has been yet made in the derivation of Eq.~(\ref{eq:eikonalequation}).

When $q_{\rm pl}$ is large enough, the exponential in the integrand on the left-hand side of Eq.~(\ref{eq:eikonalequation}) oscillates rapidly. In this case, only a small range of values of $x'$ (those for which $q_{\rm pl}|x-x'|\ll 1$) contributes to the integral and we can approximate $S(x')-S(x)$ with $[dS(x)/dx](x'-x)$. 
The left-hand side of Eq.~(\ref{eq:eikonalequation}) can therefore be estimated as
\begin{equation}\label{eq:lhs}
\simeq \left[v\left(q_{\rm pl}\sqrt{\cos^2 \theta S'(x)+\sin^2 \theta},\omega\right)\right]^{-1}~,
\end{equation}
where $S'(x)\equiv dS(x)/dx$. The right-hand side of (\ref{eq:eikonalequation}) is instead approximated with its leading order in the limit $q_{\rm pl}\to \infty$, reducing to
\begin{equation}\label{eq:rhs}
\simeq q_{\rm pl}^2 D(x)\{\sin^2 \theta +\cos^2 \theta [S'(x)]^2\}~.
\end{equation}
With these approximations, Eq.~(\ref{eq:eikonalequation}) becomes
\begin{equation}\label{eq:wkbapprox}
q_{\rm pl}\sqrt{\cos^2 \theta [S'(x)]^2+\sin^2 \theta}=q_{\rm pl}(x)~,
\end{equation}
where the {\it local} plasmon wavevector~\cite{ishmukhametov_pssb_1971,ishmukhametov_FMM_1975}  $q_{\rm pl}(x)$ is defined as the solution of
\begin{equation}
\frac{q^2_{\rm pl}(x)v(q_{\rm pl}(x),\omega)D(x)}{\pi e^2 \omega^2}=1~.
\end{equation}
The corresponding solution for the potential is, up to a multiplicative constant,
\begin{equation}\label{eq:eikonalsolution}
V_{\rm sc}(x,\theta,\omega) \propto \exp\left[i\int_{0}^x dx'\sqrt{ q^2_{\rm pl}(x')-q_{\rm pl}\sin^2 \theta }\right]~.
\end{equation}
The phase of the transmission coefficient $t_{\theta,\omega}$ is found by looking at the difference between the the solution (\ref{eq:eikonalsolution}) and the unperturbed wave, i.e.~$\exp(iq_{\rm pl}\cos \theta)$:
\begin{equation}\label{eq:argtwkb}
\arg(t_{\theta,\omega})=\int_{-\infty}^{\infty} dx' \left[ \sqrt{ q^2_{\rm pl}(x')-q_{\rm pl}\sin^2 \theta }-q_{\rm pl}\cos \theta\right]~.
\end{equation}
\section{The Method of Partial Waves}
\label{sect:partialwaves}
In this Section we introduce a decomposition of the scattering kernel and transition function in their angular components. This can be useful to treat problems with rotationally-invariant scatterers, or problems in which only a few angular components of the scattering amplitude matter.

We Fourier-decompose the incoming wave, the screened potential, the scattering kernel, and the transition function with respect to the polar angles of the relevant wavevectors:
\begin{equation}
V_{m}^{(0)}(q,\omega) =\int_{-\pi}^{\pi} \frac{d\theta_{\bm q}}{2\pi} e^{-im\theta_{\bm q}} V^{(0)}(\bm q,\omega)~,
\end{equation}
\begin{equation}
V_{{\rm sc}, m}(q,\omega) =\int_{-\pi}^{\pi}\frac{d\theta_{\bm q}}{2\pi} e^{-im\theta_{\bm q}} V_{\rm sc}(\bm q,\omega)~,
\end{equation}
\begin{eqnarray}\label{eq:kernelangular}
\Delta_{m m'}(q,q',\omega) &=&\int_{-\pi}^{\pi}\frac{d\theta_{\bm q}}{2\pi}\int_{-\pi}^{\pi}\frac{d\theta_{\bm q'}}{2\pi}e^{-im\theta_{\bm q}+im'\theta_{\bm q'}}\\ \nonumber &\times&  \Delta(\bm q,\bm q',\omega)~,
\end{eqnarray}
and
\begin{equation}
T_{mm'}(q,\omega) =\int_{-\pi}^{\pi}\frac{d\theta_{\bm q}}{2 \pi}\int_{-\pi}^{\pi}\frac{d\theta}{2 \pi}e^{-i m \theta_{\bm q} + i m' \theta}T(\bm q,\theta,\omega)~.
\end{equation}
With these definitions, Eq.~(\ref{eq:lippmannschwinger}) can be written as
\begin{equation}
\begin{split}
V_{{\rm sc}, m}(q,\omega)=V_m^{(0)}(q,\omega)+W_{\rm h}^{(+)}(q,\omega)\\
\sum_{m'=-\infty}^{\infty}\int_0^\infty \frac{dq'}{2\pi} q' \Delta_{m m'}(q,q',\omega)V_{{\rm sc}, m'}(q',\omega)~,
\end{split}
\end{equation}
while Eq.~(\ref{eq:tmatrixequation}) becomes
\begin{equation}\label{eq:tmatrixeqangular}
\begin{split}
& T_{mm'}(q,\omega)= \Delta_{m m'}(q,q_{\rm pl},\omega)\\
& +\sum_{n=-\infty}^{\infty} \int_0^\infty \frac{dq'}{2\pi} q'\Delta_{m n}(q,q',\omega)W^{(+)}_{\rm h}(q',\omega)T_{nm'}(q',\omega)~.
\end{split}
\end{equation}
In Section~\ref{sect:example} we present the results of a numerical solution this equation in a concrete situation.
 
Once Eq.~(\ref{eq:tmatrixeqangular}) is solved, the scattering amplitude can be easily calculated by using
\begin{equation}
T(\bm q,\theta,\omega)=\sum_{m=-\infty}^\infty \sum_{m'=-\infty}^\infty e^{i m \theta_{\bm q}-im'\theta}T_{mm'}(q,\omega)~.
\end{equation}
Even in this geometry, the presence of a localized plasmon resonanc manifests as a pole of the analytical continuation of $T_{mm'}(q,\theta,\omega)$ to the upper half of the complex plane, at a purely imaginary value of $q$.

For systems with rotational invariance, only the diagonal components of $\Delta_{mm'}$ and $T_{mm'}$ are non-zero. This greatly simplifies the solution of Eq.~(\ref{eq:tmatrixeqangular}). 
In this case, the scattering amplitude depends only on the angle between the incoming and scattered waves and can be written as
\begin{equation}\label{eq:frotationalinvariant}
\begin{split}
f(\theta,\omega)  & \equiv f(\theta,0,\omega)\\
&=-e^{i\frac{\pi}{4}}\frac{\sqrt{q_{\rm pl}}C(\omega)}{\sqrt{2\pi}}\sum_{m=-\infty}^{+\infty}e^{im\theta}T_{m}(q_{\rm pl},\omega)~,
\end{split}
\end{equation}
where $T_m(q,\omega)\equiv T_{mm}(q,\omega)$. Using Eq.~(\ref{eq:frotationalinvariant}) and the optical theorem (\ref{eq:opticaltheorem}) we obtain 
\begin{eqnarray}\label{eq:transitionmatrixconstraint}
&&\sum_{m=-\infty}^\infty \frac{|q_{\rm pl}C(\omega)T_m(q_{\rm pl},\omega)|^2}{2} \nonumber\\
&+&\Im m[q_{\rm pl}C(\omega)T_m(q_{\rm pl},\omega)]=0~.
\end{eqnarray}
Since the quantities $T_{m}(q_{\rm pl},\omega)$ for different values of $m$ are independent from each other, every term of the sum in Eq.~(\ref{eq:transitionmatrixconstraint}) must vanish. This restricts the region of the complex plane allowed for the values of $q_{\rm pl}C(\omega)T_m(q_{\rm pl},\omega)$ to a circle of radius $1$ centered in $-i$. This region can be parametrized by a single real number $-\pi/2 \leq\delta_{m,\omega}\leq \pi/2$, called phase shift, in the following way
\begin{equation}\label{eq:transitionmatrixparametrization}
q_{\rm pl}C(\omega)T_m(q_{\rm pl},\omega)=-2\sin(\delta_{m,\omega})e^{i\delta_{m,\omega}}~.
\end{equation}
We can therefore express the scattering amplitude and the total cross section in terms of the phase shifts in a compact way:
\begin{equation}\label{eq:amplitudephaseshifts}
f(\theta,\omega)=\frac{2e^{i\pi/4}}{\sqrt{2\pi q_{\rm pl}}}\sum_{m=-\infty}^\infty\sin(\delta_{m,\omega})e^{i\delta_{m,\omega}+im\theta}
\end{equation}
and
\begin{equation}\label{eq:crosssectionphaseshifts}
\Sigma(\omega)\equiv \int d\theta |f(\theta,\omega)|^2 =\frac{4}{ q_{\rm pl}}\sum_{m=-\infty}^\infty\sin^2(\delta_{m,\omega})~.
\end{equation}
\section{Explicit examples}
\label{sect:example}
In this Section we illustrate the power of our Lippmann-Schwinger theory by solving two concrete problems, one for each of the geometries displayed in Fig.~\ref{fig:geometry}.

We consider the scattering of plasmons in a 2D parabolic-band electron gas subject to an external scalar perturbation generated by: (a) a charged point-like impurity with charge $Ze$, $Z$ being an integer number, positioned at $(x,y,z)=(0,0,d)$ and (b) a line of charged impurities with charge density per unit length $\lambda$, 
positioned at $x=0$ and $z=d$. Here, $z=0$ is the position of the 2D electron gas, which, in the absence of the impurities, has a uniform density $\bar{n}$. 
 For the sake of definiteness, we take $\bar{\epsilon}=12$ and $m = 0.067~m_{\rm e}$, where $m_{\rm e}$ is the electron mass in vacuum. These material parameters refer to a 2D parabolic-band electron gas in a GaAs quantum well.

\begin{figure}[h!!]
\begin{overpic}[width=.98\linewidth]{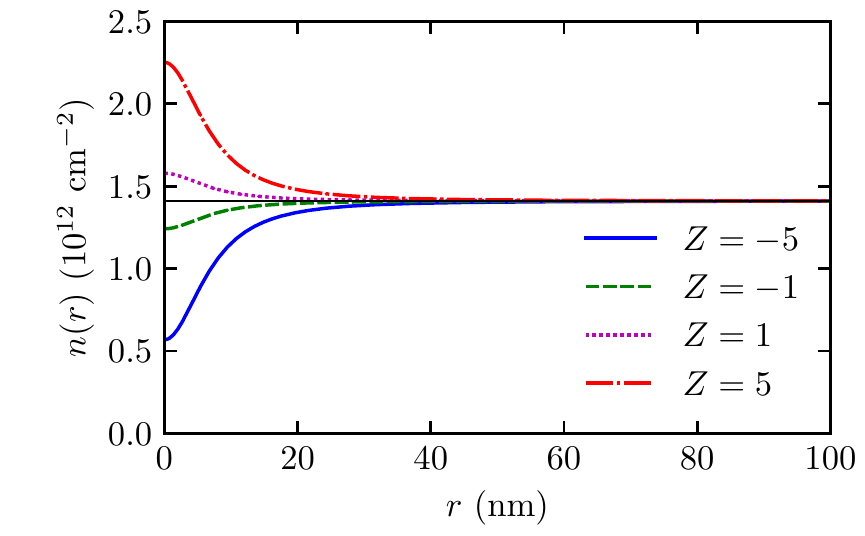}\put(2,150){(a)}\end{overpic}
\begin{overpic}[width=.98\linewidth]{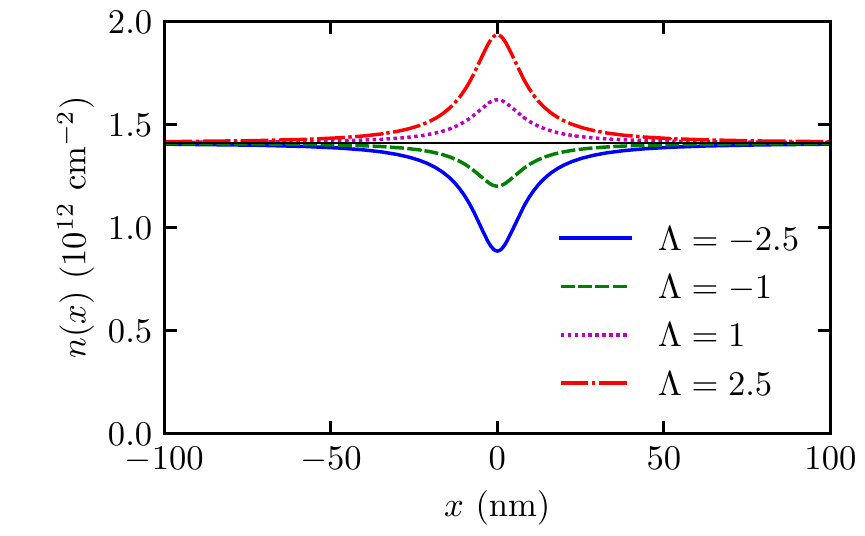}\put(2,150){(b)}\end{overpic}
\begin{overpic}[width=.98\linewidth]{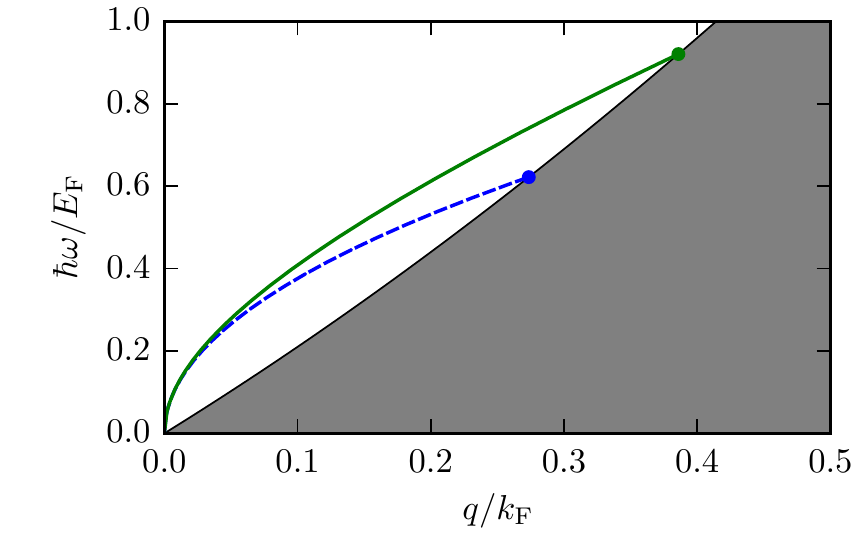}\put(2,150){(c)}\end{overpic}
\caption{\label{fig:dispersion} (Color online) Panel (a) Spatial dependence $n(r)$ of the density profile induced by a point-like charged impurity located above a 2D parabolic-band electron gas, at $(x,y)=(0,0)$ and $z=d=2/k_{\rm F}$. These results have been obtained for a Wigner-Seitz density parameter~\cite{Giuliani_and_Vignale} $r_{\rm s}=0.5$. (It is only in this weak-coupling limit that the application of RPA is rigorously justified~\cite{Giuliani_and_Vignale}.) Different curves refer to different values of the impurity charge $Ze$. Panel (b) Same as in panel (a) but for a line of charged impurities located at $x=0$ and $z=d=2/k_{\rm F}$. Different curves refer to different values of the dimensionless parameter $\Lambda$ introduced in Eq.~(\ref{eq:coupling_constant}). Panel (c) RPA dispersion relation of plasmons in a uniform 2D parabolic-band electron system (evaluated at $r_{\rm s}=0.5$). The blue dashed line is the result of the local theory, obtained by setting $\mathcal{G}(q,\omega)=1$, while the green solid line is the result of our nonlocal theory. The grey-shaded area represents the electron-hole continuum, where plasmons suffer Landau damping~\cite{Giuliani_and_Vignale}.}
\end{figure}

The electric potential generated by the external charges perturbs the uniform ground-state density inducing a non-trivial density profile $n(\bm r) = \bar{n} +\delta n(\bm r)$, which depends only on $r=|\bm r|$ in geometry (a) and only on $x$ in geometry (b). These density profiles are shown in Fig.~\ref{fig:dispersion}(a) and~(b). Details on how $n(\bm r)$ is actually calculated are reported below.

As explained in Sect.~\ref{sect:scatteringproblem}, 
we first need to calculate the homogeneous part $\tilde{\chi}_{\rm h}(q,\omega)$ 
of the density-density response function of the system in the absence of the perturbations, i.e.~for $Z=0$ in geometry (a) and for $\lambda=0$ in geometry (b). 
To this end, we use Eqs.~(\ref{eq:powerexpansion})-(\ref{eq:moments}) with $\bm q' = \bm q$, retaining terms ${\cal O}(\omega^{-4})$ (i.e.~expanding up to $\ell=3$). The first moment is given by
\begin{equation}
M^{(1)}(\bm q, \bm q) = \frac{\bar{n} }{m}q^2~, 
\end{equation}
the second moment $M^{(2)}(\bm q, \bm q)$ is identically zero, while the third moment reads as following:
\begin{equation}
M^{(3)}(\bm q, \bm q) = \frac{3 \bar{n}\epsilon_{0}}{m^2}q^4
 + \frac{\bar{n} \hbar^2}{4 m^3} q^6 ~,
\end{equation}
where $\epsilon_{0} = {E_{\rm F}}/2$ is the kinetic energy per particle of the non-interacting 2D electron system~\cite{Giuliani_and_Vignale}. Here, $E_{\rm F} = \pi \bar{n} \hbar^2/m$ is the Fermi energy.
Using these three results we find that $\tilde{\chi}_{\rm h}(q,\omega)$ can be expressed as in 
Eq.~(\ref{eq:chih}) with $D=\pi e^2 \bar{n}/m$ and 
\begin{equation}
{\cal G}(q,\omega)=1+\frac{3}{4}\frac{v_{\rm F}^2 q^2}{\omega^2}+\frac{1}{4}\frac{\hbar^2q^4}{m^2 \omega^2}~,
\end{equation}
where $v_{\rm F} = \hbar k_{\rm F}/m$ is the Fermi velocity and $k_{\rm F} = \sqrt{2\pi \bar{n}}$ is the Fermi wave number.

For the sake of simplicity, in Eq.~(\ref{eq:vh}) we neglect the frequency dependence of $\bar{\epsilon}(\omega)$, by taking $\bar{\epsilon}(\omega) \mapsto \bar{\epsilon}$, and also finite-size effects, by setting ${\cal F}(q,\omega)\equiv 1$. Using Eq.~(\ref{eq:resdefinition}) we find
\begin{equation}\label{eq:C}
C(\omega)=\frac{2\pi e^2}{\bar{\epsilon}} \frac{1+\frac{3}{4}\frac{v_{\rm F}^2 q^2_{\rm pl}}{\omega^2}+\frac{1}{4}\frac{\hbar^2q^4_{\rm pl}}{m^2 \omega^2}}{1+\frac{9}{4}\frac{v_{\rm F}^2 q^2_{\rm pl}}{\omega^2}+\frac{5}{4}\frac{\hbar^2q^4_{\rm pl}}{m^2 \omega^2}}~,
\end{equation}
while using Eq.~(\ref{eq:Wcorrection}) we obtain the nonlocal correction to the effective interaction
\begin{equation}\label{eq:wcorr}
\begin{split}
& {\cal W}(q,\omega)=\frac{q_{\rm pl}-q}{q}\frac{E_{\rm F}^2q_{\rm pl}^2}{\hbar^2\omega^2k_{\rm F}^2}\times\\
& \frac{3\left(\frac{q}{q_{\rm pl}}+2\right)+\frac{q_{\rm pl}^2}{k_{\rm F}^2}\left(\frac{q^3}{q_{\rm pl}^3}+\frac{2q^2}{q_{\rm pl}^2}+\frac{3q}{q_{\rm pl}}+4\right)}{\left(1-\frac{q}{q_{\rm pl}}\right)+\frac{3E_{\rm F}^2q_{\rm pl}^2}{\hbar^2\omega^2k_{\rm F}^2}\left(1-\frac{q^3}{q_{\rm pl}^3}\right)+\frac{E_{\rm F}^2q_{\rm pl}^4}{\hbar^2\omega^2k_{\rm F}^4}\left(1-\frac{q^5}{q_{\rm pl}^5}\right)}~.
\end{split}
\end{equation}
Plasmon modes of the uniform 2D parabolic-band electron system analyzed in this work are shown in Fig.~\ref{fig:dispersion}(c).
Results of the local theory (i.e.~obtained by negleting $M^{(3)}(\bm q,\bm q')$) are simply $\mathcal{G}(q,\omega)=1$, $C(\omega)=2\pi e^2/\bar{\epsilon}$ and $\mathcal{W}(q,\omega)=0$.
We now analyze separately the two geometries (a) and (b). In case of geometry (a), we calculate the phase shifts $\delta_{m,\omega}$ and the scattering cross section $\Sigma(\omega)$ as functions of the plasmon wavevector and impurity charge $Z$, limiting ourselves to the local approximation. In the case of geometry (b), we calculate transmission and reflection coefficients as functions of the plasmon wavevector and impurity charge density $\lambda$ using the full nonlocal theory and compare these results with the corresponding ones in the local approximation.

\subsection{Scattering of a plasmon against a point-like charged impurity}
The potential generated by a charge $eZ$ located at a distance $d$ from the plane of the 2D electron gas is
\begin{equation}
U_{\rm ext}(r)=-\frac{e^2Z}{\bar{\epsilon}\sqrt{r^2+d^2}}~.
\end{equation}
Its Fourier transform reads as following
\begin{equation}
U_{\rm ext}(q)=-\frac{2\pi e^2Ze^{-qd}}{\bar{\epsilon}q}~.
\end{equation}
We can calculate the density-density response function of the non-uniform system using the results in Appendix~\ref{sect:chi}.
Retaining only the first moment $M^{(1)}(\bm q,\bm q')$ we obtain
\begin{equation}\label{eq:kernel0d}
\Delta(\bm q,\bm q',\omega)=\delta\tilde{\chi}(\bm q,\bm q',\omega)=\frac{\bm q\cdot\bm q' }{m\omega^2}\delta n(|\bm q-\bm q'|)~.
\end{equation}
In this approximation the scattering kernel depends only on the induced density perturbation.
We now evaluate $\delta n(q)$ by using linear response theory~\cite{Giuliani_and_Vignale} with respect to $U_{\rm ext}(r)$ and the RPA.
The total potential is
\begin{equation}
U_{\rm tot}(q) =\frac{U_{\rm ext}(q)}{\epsilon(q)}~,
\end{equation}
where the static dielectric constant of the uniform 2D parabolic-band electron gas is~\cite{Giuliani_and_Vignale,Stern2DEG}
\begin{eqnarray}\label{eq:epsilon_Stern}
\epsilon(q) &=& 1- \frac{2\pi e^2}{\bar{\epsilon}q} \chi_0(q,\omega=0) \nonumber\\
&=& 1+\frac{q_{\rm TF}}{q}\left[1-\Theta(q-2k_{\rm F})\frac{\sqrt{q^2-4k_{\rm F}^2 }}{q}\right]~.
\end{eqnarray}
Here, $\chi_0(q,\omega=0)$ is the static density-density response function of a 2D parabolic-band electron gas~\cite{Giuliani_and_Vignale,Stern2DEG} and $q_{\rm TF}=2m e^2/(\hbar^2\bar{\epsilon})$ is the Thomas-Fermi wave number~\cite{Giuliani_and_Vignale,Stern2DEG}.

Making use of well-known analytical expressions~\cite{Giuliani_and_Vignale,Stern2DEG} for $\chi_0(q,\omega=0)$, we can write the density perturbation as
\begin{equation}\label{eq:deltan0d}
\delta n(q)= - N_{0} \left[1-\Theta(q-2k_{\rm F})\frac{\sqrt{q^2-4k_{\rm F}^2 }}{q}\right]U_{\rm tot}(q)~,
\end{equation}
where $N_{0}=m/(\pi\hbar^2)$ is the density of states at the Fermi energy~\cite{Giuliani_and_Vignale}.
We now make a further approximation neglecting all the terms that are proportional to $\Theta(|q_x|-2k_{\rm F})$. Indeed, since the plasmon wavevector $q_{\rm pl}$ is a fraction of $k_{\rm F}$, we expect that the contribution to the scattering problem coming from wavevectors satisfying $|q_{x}|> 2k_{\rm F}$ is negligible. This amounts to neglecting Friedel oscillations of the electron density. The total density profile in real space $n(r) = \bar{n} + \delta n(r)$ is shown Fig.~\ref{fig:dispersion}(a).

\begin{figure}
\begin{overpic}[width=\linewidth]{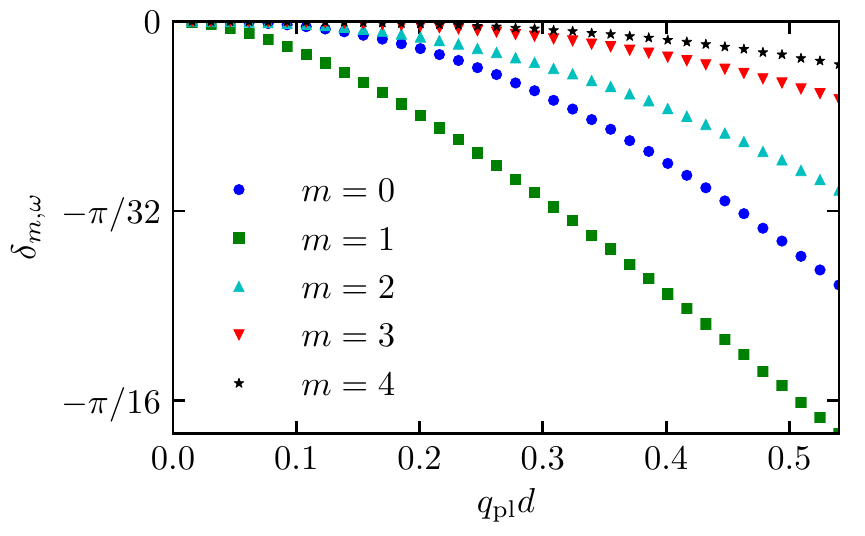}\put(2,150){(a)}\end{overpic}
\begin{overpic}[width=\linewidth]{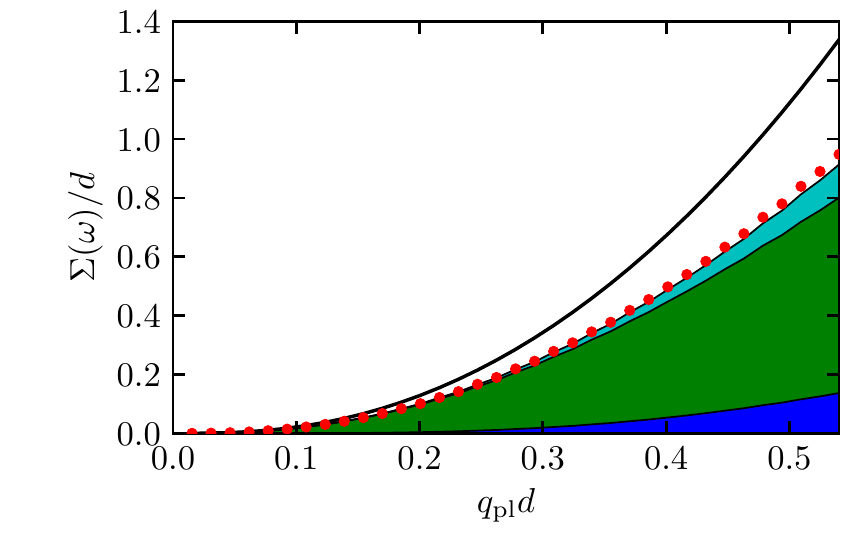}\put(2,150){(b)}\end{overpic}
\caption{\label{fig:pointchargeqddependence} (Color online) Panel (a) The numerically calculated phase shifts $\delta_{m,\omega}$ for the case of geometry (a) are plotted as functions of the plasmon wavevector $q_{\rm pl}$ for $Z=5$ and $0\leq m \leq 4$.  Panel (b) The total cross section $\Sigma(\omega)$ (calculated by including in the numerics partial waves with $|m|\leq 5$) is plotted as a function of the plasmon wavevector $q_{\rm pl}$ for $Z=5$ (red dots). The thick black line is the result of first-order Born approximation. The blue, green, and cyan-shaded regions denote the contributions to the total cross section of partial waves with $m=0$, $m=\pm 1$, and $m=\pm 2$, respectively. Note that the dominant contribution comes from the $m=\pm 1$ channel ($p$-wave scattering).}
\end{figure}

\begin{figure}
\begin{overpic}[width=\linewidth]{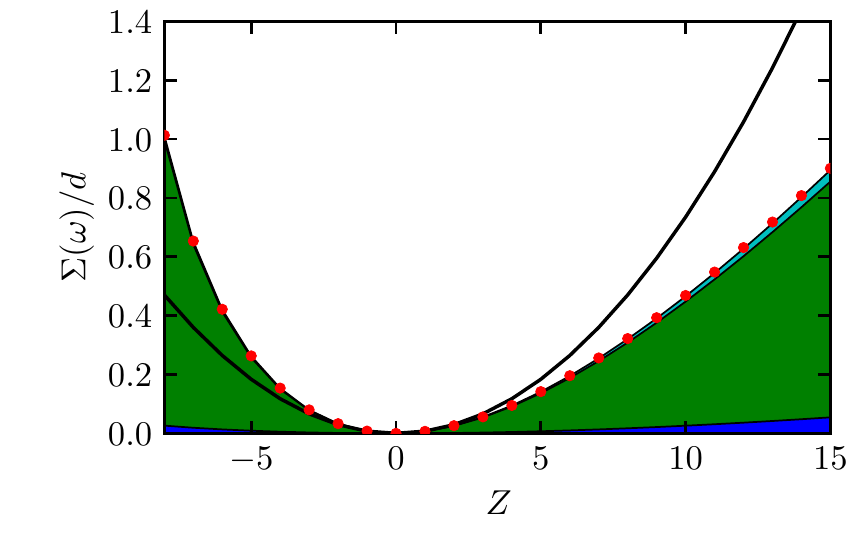}\put(2,150){(a)}\end{overpic}
\begin{overpic}[width=\linewidth]{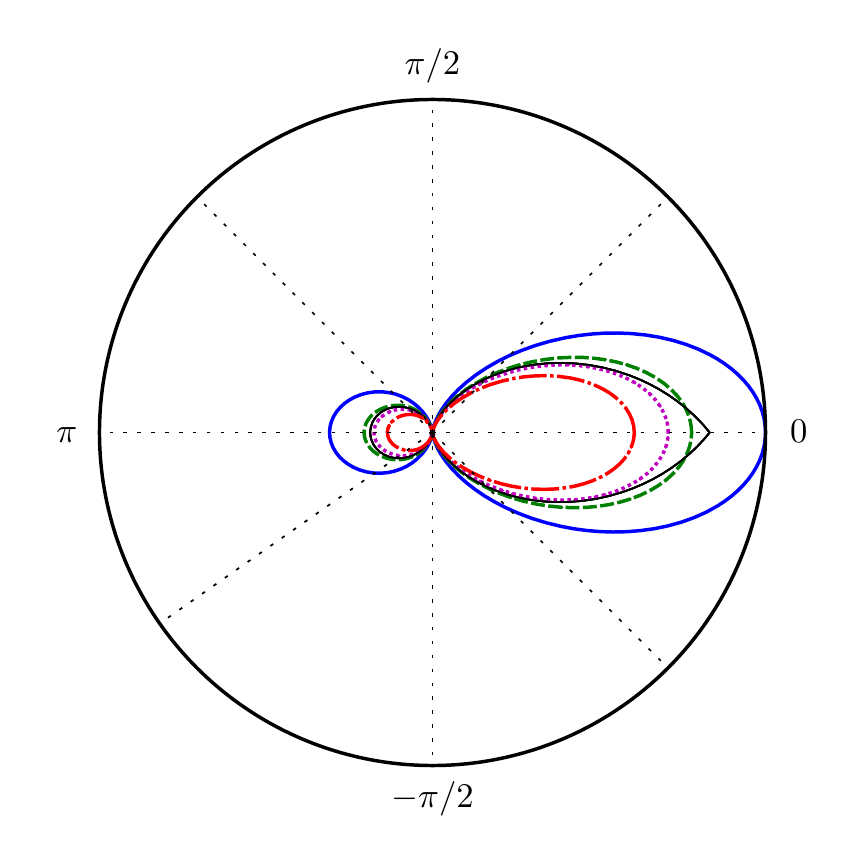}\put(2,240){(b)}\end{overpic}
\caption{\label{fig:pointchargeZdependence} (Color online) Panel (a) Total scattering cross section $\Sigma(\omega)$ as a function of the impurity charge $Z$, for a fixed value of the plasmon wavevector, i.e.~ $q_{\rm pl}d=0.23$ (red dots). The thick black line is the result of the first-order Born approximation. The blue, green, and cyan-shaded regions denote the contributions to the total cross section of partial waves with $m=0$, $m=\pm 1$, and $m=\pm 2$, respectively. Panel (b) Angular dependence of the numerically evaluated  scattering cross section normalized by $Z^2$, $|f(\theta)|^2/(Z^{2} d)$, for $q_{\rm pl}d=0.23$ and different values of $Z$: $Z=-5$ (blue solid line), $Z=-1$ (green dashed line), $Z=1$ (magenta dotted line), and $Z=5$ (red dash-dotted line). The black line is the result of the first-order Born approximation. Again, only partial waves with $|m|\leq 5$ were kept in all numerical calculations.}
\end{figure}

Using Eqs.~(\ref{eq:kernel0d})-(\ref{eq:deltan0d}) we can calculate the dimensionless scattering kernel, obtaining the following expression
\begin{equation}\label{eq:kernel20d}
q_{\rm pl}C(\omega)\Delta(\bm q,\bm q',\omega)=\frac{2\pi e^2Z}{\bar{\epsilon}E_{\rm F}}\frac{\bm q \cdot \bm q' e^{-d|\bm q-\bm q'|}}{|\bm q-\bm q'|+q_{\rm TF}}~.
\end{equation}
This quantity is related by Eq.~(\ref{eq:FOBA}) to the scattering amplitude in the first-order Born approximation:
\begin{equation}
f^{(1)}(\theta,\omega)=- \frac{e^{i\frac{\pi}4}q_{\rm pl}^{3/2}}{\sqrt{2\pi }}\frac{2\pi e^2Z}{\bar{\epsilon}E_{\rm F}}\frac{\cos(\theta) e^{-d q_{\rm pl}\sqrt{2[1-\cos(\theta)]}}}{q_{\rm pl}\sqrt{2[1-\cos(\theta)]}+q_{\rm TF}}~.
\end{equation}
This is the most important analytical result of this Section.
Note the presence of the overall factor $\cos{(\theta)}$, which is responsible for the dominance of $p$-wave (i.e.~$m=\pm 1$) scattering---see Fig.~\ref{fig:pointchargeqddependence}(b)--- and for the suppression of scattering in the direction perpendicular to the incident one---see Fig.~\ref{fig:pointchargeZdependence}(b).


Since the problem at hand is rotationally symmetric, we can decompose Eq.~(\ref{eq:tmatrixequation}) for the transition function into its cylindrical components, as described in Sect.~\ref{sect:partialwaves}. This procedure requires the knowledge of the angular components of the scattering kernel defined in Eq.~(\ref{eq:kernelangular}):
\begin{equation}
\begin{split}
& q_{\rm pl}C(\omega)\Delta_{mm}(q, q',\omega)=\\
& \frac{2\pi e^2Z}{\bar{\epsilon}E_{\rm F}}q q'\int_{-\pi}^\pi \frac{d\theta}{2\pi} e^{-im\theta}\frac{\cos(\theta) e^{-d\sqrt{q^2+q'^2-2qq'\cos(\theta)}}}{\sqrt{q^2+q'^2-2qq'\cos(\theta)}+q_{\rm TF}}~.
\end{split}
\end{equation}
This quantity needs to be evaluated numerically for each $m$.

\subsection{Scattering of a plasmon against a 1D line of charged impurities}

The external potential generated by a line of charges can be calculated via Gauss theorem and reads
\begin{equation}
U_{\rm ext}(x)=\frac{e\lambda}{\bar{\epsilon}} \ln \left(1+\frac{x^2}{d^2}\right)~.
\end{equation}
Its Fourier transform is
\begin{equation}
U_{\rm ext}(q_x)=-\frac{2\pi e \lambda e^{-d|q_x|}}{\bar{\epsilon} |q_x|}~.
\end{equation}
We are clearly in the case of Fig.~\ref{fig:geometry}(b), with translational invariance along the $\hat{\bm y}$ direction.

We are now in the position to calculate the scattering kernel. Using Eq.~(\ref{eq:Udefinition}) and the results 
(\ref{eq:M1}), (\ref{eq:M2}), and~(\ref{eq:M3}) we find:
\begin{equation}\label{eq:kernel}
\begin{split}
& \Delta(q_x,q_x',q_y,\omega)=\frac{(q_x q_x'+q_y^2)}{m\omega^2}\delta n(q_x-q_x')\\
&+\frac{3(q_xq_x'+q_y^2)}{m^2 \omega^4}\left[q_xq_x'\delta T_{xx}(q_x-q_x')+q_y^2 \delta T_{yy}(q_x-q_x')\right]\\
& +\frac{\hbar^2}{m^3\omega^4}(q_xq_x'+q_y^2)\left[\frac{3}{4}q_xq_x'(q_x-q_x')^2+\frac{1}{4}(q_x q_x'+q_y^2)^2\right]\\
&\times\delta n(q_x-q_x')+\frac{1}{m\omega^4}q_xq_x'\left\langle\partial_{xx} U_{\rm tot} n_{\hat{\bm x}(q_x-q_x')}\right\rangle~.
\end{split}
\end{equation}
Here $\delta n(q_x)$ is the variation of the electron density with respect to its equilibrium value $\bar{n}$, while $\delta T_{ij}(q_x)$ represents the variations of the stress-tensor components with respect to their the equilibrium values.

To get an explicit analytical expression for the scattering kernel, 
we evaluate the expectation values in Eq.~(\ref{eq:kernel}) 
by using linear response theory~\cite{Giuliani_and_Vignale} with respect to $U_{\rm ext}(x)$ and the RPA. We follow the same steps as in the previous Section. We start again by calculating the Fourier transform of the total potential $U_{\rm tot}(x)$:
\begin{equation}
U_{\rm tot}(q_x) =\frac{U_{\rm ext}(q_x)}{\epsilon(|q_x|)}~.
\end{equation}
The density perturbation reads as following
\begin{equation}
\delta n(q_x)= - N_{0} \left[1-\Theta(|q_x|-2k_{\rm F})\frac{\sqrt{q_x^2-4k_{\rm F}^2 }}{|q_x|}\right]U_{\rm tot}(q_x)~.
\end{equation}
The resulting inverse Fourier transform of the density profile $n(x) = \bar{n} + \delta n(x)$ is plotted as a function of $x$ in Fig.~\ref{fig:dispersion}(b).

The expectation value of the second derivative of the potential is, to linear order in $U_{\rm ext}$,
\begin{equation}
\left\langle\partial_{xx}U_{\rm tot}(x)n_{\hat{\bm x} q_x}\right\rangle=-q_x^2\bar{n}U_{\rm tot}(q_x)~.
\end{equation}
The components of the stress tensor can be evaluated using the density-stress tensor response function calculated in Appendix \ref{app:stress}. We find
\begin{equation}
\delta T_{xx}(q_x)=-N_0E_{\rm F}f_x(|q_x|)U_{\rm tot}(q_x)
\end{equation}
and
\begin{equation}
\delta T_{yy}(q_x)=-N_0E_{\rm F}f_y\left(\left|q_x\right|\right)U_{\rm tot}(q_x)~.
\end{equation}
Here,
\begin{equation}\label{eq:fxdefinition}
f_x(q)=1+\frac{q^2}{2k_{\rm F}^2}-\Theta(q-2k_{\rm F})\frac{q}{2k_{\rm F}^2}\sqrt{q^2-4k_{\rm F}^2 }
\end{equation}
and
\begin{equation}\label{eq:fydefinition}
f_y(q)=1-\frac{q^2}{6k_{\rm F}^2}-\Theta(q-2k_{\rm F})\frac{2\sqrt{q^2-4k_{\rm F}^2}}{3q}\left(1-\frac{q^2}{4k_{\rm F}^2}\right)~.
\end{equation}

We neglect again all the terms in the expectation values that are proportional to $\Theta(|q_x|-2k_{\rm F})$. 
In summary, our final result for the dimensionless scattering kernel is
\begin{equation}\label{eq:kernel2}
\begin{split}
& C(\omega)\Delta(q_x,q_x',q_y,\omega)=\\
&=\frac{\Lambda e^{-d|q_x-q_x'|}(q_xq_x'+q_{\rm pl}^2\sin^2 \theta)}{q_{\rm pl}\left(|q_x-q_x'|+q_{\rm TF}\right)}{\cal K}(q_x,q_x',q_y,\omega)~,
\end{split}
\end{equation}
where
\begin{equation}\label{eq:coupling_constant}
\Lambda = \frac{2\pi e\lambda}{\bar{\epsilon} E_{\rm F}}
\end{equation}
is the dimensionless ``impurity'' concentration (for $\Lambda> 0$ the external potential is attractive for the electron system, while for $\Lambda<0$ it is repulsive) and ${\cal K}(q_x,q_x',q_y,\omega)$ is a function that takes into account nonlocal effects: 
\begin{widetext}
\begin{equation}\label{eq:Kcorrection}
\begin{split}
& {\cal K}(q_x,q_x',q_y,\omega)=
 \frac{1+\frac{2E_{\rm F}^2}{\hbar^2\omega^2k_{\rm F}^2}\left[3(q_xq_x'+q_y^2 )+\frac{q_xq_x'(q_x-q_x')^2}{q_xq_x'+q_y^2}\right]
+\frac{E_{\rm F}^2}{\hbar^2\omega^2k_{\rm F}^4}\left[(6q_xq_x'-q_{y}^2)(q_x-q_x')^2+(q_xq_x'+q_{y}^2 )^2\right]}
{1+\frac{9E_{\rm F}^2q_{\rm pl}^2}{\hbar^2\omega^2k_{\rm F}^2}+\frac{5 E_{\rm F}^2q_{\rm pl}^4}{\hbar^2\omega^2k_{\rm F}^4}}~.
\end{split}
\end{equation}
\end{widetext}
We can now make use of Eqs.~(\ref{eq:tFOBA})-(\ref{eq:rFOBA}) to evaluate the transmission and reflection coefficients in the first-order Born approximation. We find
\begin{equation}\label{eq:tFOBAline}
t_{\theta, \omega}^{(1)}=1-i\frac{\Lambda q_{\rm pl}}{q_{\rm TF}\cos\theta}\mathcal{K}(q_{\rm pl}\cos\theta,q_{\rm pl}\cos\theta,q_{\rm pl}\sin\theta,\omega)
\end{equation}
and
\begin{equation}\label{eq:rFOBAline}
\begin{split}
r_{\theta, \omega}^{(1)}=& i\frac{\Lambda q_{\rm pl} \cos(2\theta)e^{-2dq_{\rm pl}\cos \theta}}{\cos\theta(2q_{\rm pl}\cos \theta+q_{\rm TF})}\times\\
& \mathcal{K}(-q_{\rm pl}\cos\theta,q_{\rm pl}\cos\theta,q_{\rm pl}\sin\theta,\omega)~.
\end{split}
\end{equation}
\subsection{Numerical results}

In the case of geometry (a),  we solved numerically Eq.~(\ref{eq:tmatrixeqangular}) by using a first-order finite-element method for partial waves with $0\leq m\leq 5$ and the local results for $C(\omega)$ and $\mathcal{W}(q,\omega)$, together with Eq.~(\ref{eq:kernel20d}). From the resulting angular components of the transition matrix we extracted the corresponding phase shifts $\delta_{m,\omega}$ by inverting Eq.~(\ref{eq:transitionmatrixparametrization}). (Results for negative values of $m$ are readily obtained by using $\delta_{m,\omega}=\delta_{-m,\omega}$.)  Numerical results for the phase shifts are shown in Fig.~\ref{fig:pointchargeqddependence}(a).
We then used the phase shifts to calculate the scattering amplitude $f(\theta,\omega)$ and the total cross section $\Sigma(\omega)$ according to Eqs.~(\ref{eq:amplitudephaseshifts})-(\ref{eq:crosssectionphaseshifts}). Numerical results for the cross section are shown in Figs.~\ref{fig:pointchargeqddependence}(b) and~\ref{fig:pointchargeZdependence}(a) and compared to the results of the first-order Born approximation. Fig.~\ref{fig:pointchargeZdependence} shows the angular distribution of the scattered power (proportional to the square modulus of the scattering amplitude) for a fixed value of the plasmon wavevector $q_{\rm pl}$ and different values of $Z$. We note that most of the power is scattered in the forward direction inside an angle of $\approx\pm 45^{\circ}$ from the incidence direction.
A smaller fraction of the power is backscattered, while ``lateral'' scattering is almost negligible. 

\begin{figure}
\begin{overpic}[width=\linewidth]{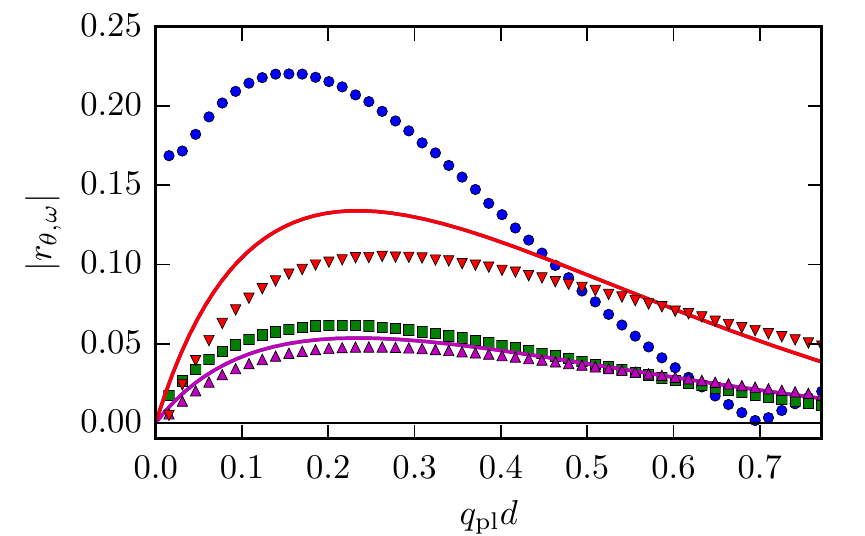}\put(2,150){(a)}\end{overpic}
\begin{overpic}[width=\linewidth]{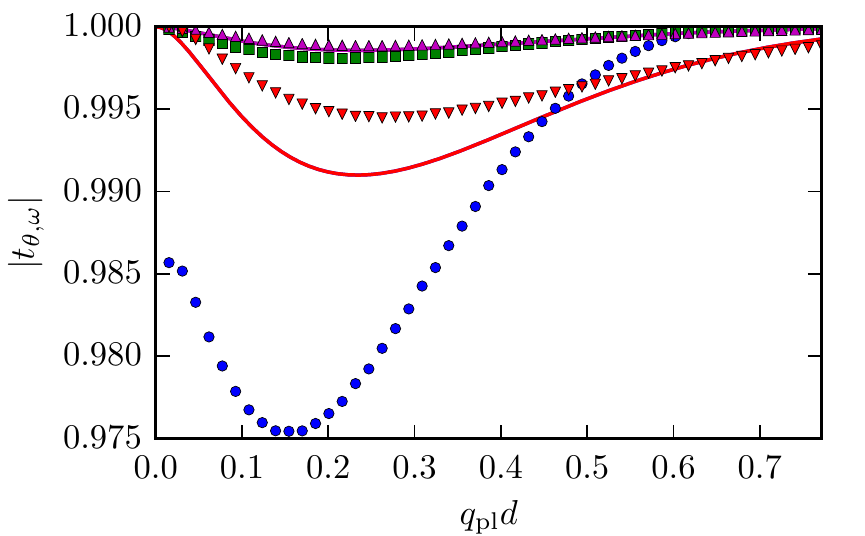}\put(2,150){(b)}\end{overpic}
\caption{\label{fig:amplitudes} (Color online) Numerically evaluated amplitudes of the reflection---panel (a)---and transmission---panel (b)---coefficients, as functions of the plasmon wavevector $q_{\rm pl}$, for different values and signs of the dimensionless parameter $\Lambda$: $\Lambda=-2.5$ (blue circles), $\Lambda=-1$ (green squares), $\Lambda=1$ (magenta upward triangles), and $\Lambda=2.5$ (red downward triangles). These results include nonlocal effects. The solid lines with the same color coding are the results of the first-order Born approximation. In this approximation the results are even in $\Lambda$, therefore only curves corresponding to positive values of $\Lambda$ are shown.}
\end{figure}

\begin{figure}
\begin{overpic}[width=\linewidth]{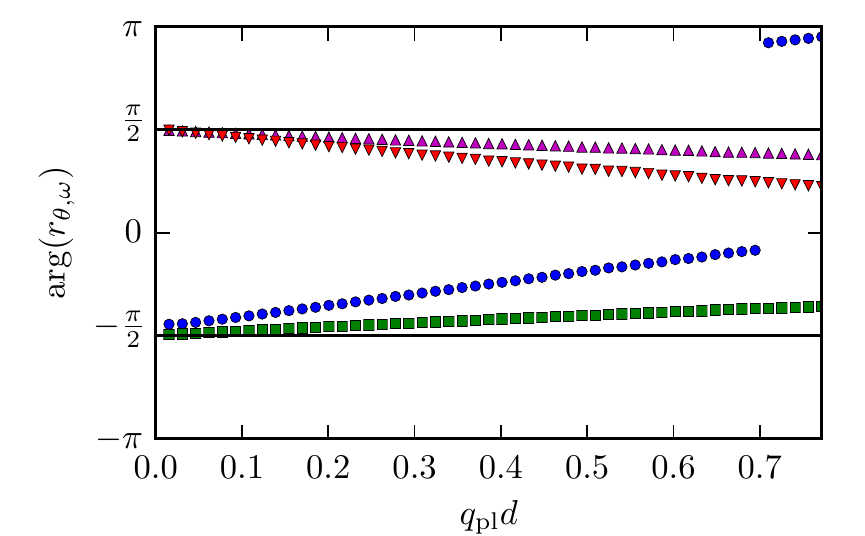}\put(2,150){(a)}\end{overpic}
\begin{overpic}[width=\linewidth]{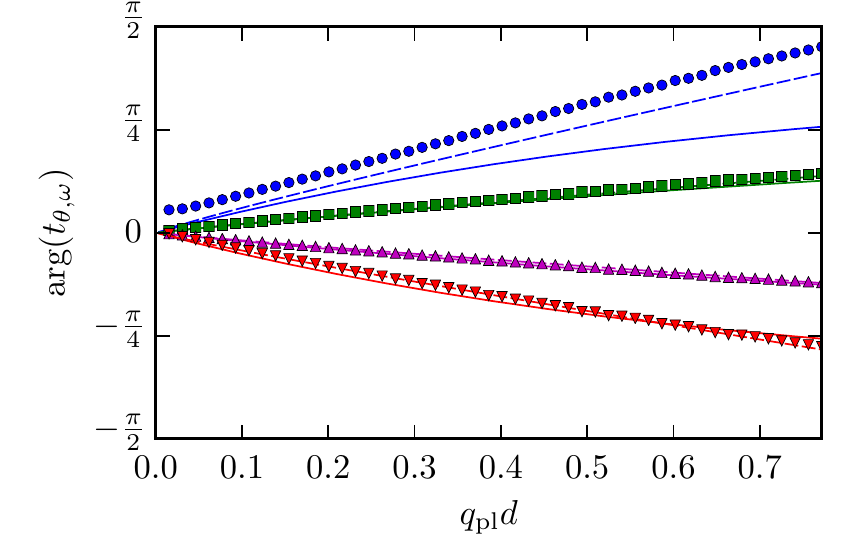}\put(2,150){(b)}\end{overpic}
\caption{\label{fig:phases} (Color online) Numerically evaluated phases of the reflection---panel (a)---and transmission---panel (b)---coefficients, as functions of the dimensionless product $q_{\rm pl}d$, for different values and signs of the dimensionless parameter $\Lambda$: $\Lambda=-2.5$ (blue circles), $\Lambda=-1$ (green squares), $\Lambda=1$ (magenta upward triangles), and $\Lambda=2.5$ (red downward triangles). The black solid lines in panel (a) represent the result of the first-order Born approximation. The solid lines in panel (b) represent the result of the first-order Born approximation, while the dashed lines are the results of the eikonal approximation.}
\end{figure}

For the case of geometry (b), we solved numerically Eq.~(\ref{eq:tmatrixequation1d}) by using a first-order finite-element method, making use of the expressions in Eqs.~(\ref{eq:C}), (\ref{eq:wcorr}), (\ref{eq:kernel2}), and~(\ref{eq:Kcorrection}).  All numerical results for $r_{\theta,\omega}$ and $t_{\theta,\omega}$ have been obtained by setting $\theta=0$ and evaluating $\omega$ at the plasmon dispersion. This implies that $\omega$ changes with $q_{\rm pl}$, as dictated by the RPA equation (\ref{eq:kpdefinition}). A summary of our main results for the transmission and reflection coefficients as functions of the plasmon wavevector $q_{\rm pl}$ is presented in Figs.~\ref{fig:amplitudes}-\ref{fig:phases}. Full numerical results (denoted by symbols) are compared with the results of the first-order Born and eikonal approximations. We clearly see that the first-order Born approximation works well for the amplitude of reflection and transmission coefficients (Fig.~\ref{fig:amplitudes}) in the weak-coupling limit $|\Lambda| \ll 1$. The same approximation works well in the same limit for the phase of the transmission coefficient, as shown in Fig.~\ref{fig:phases}(b). From the same figure, it is also clear that the eikonal approximation performs better than the first-order Born approximation in predicting $\arg(t)$, especially at strong coupling.

\begin{figure}
\begin{overpic}[width=\linewidth]{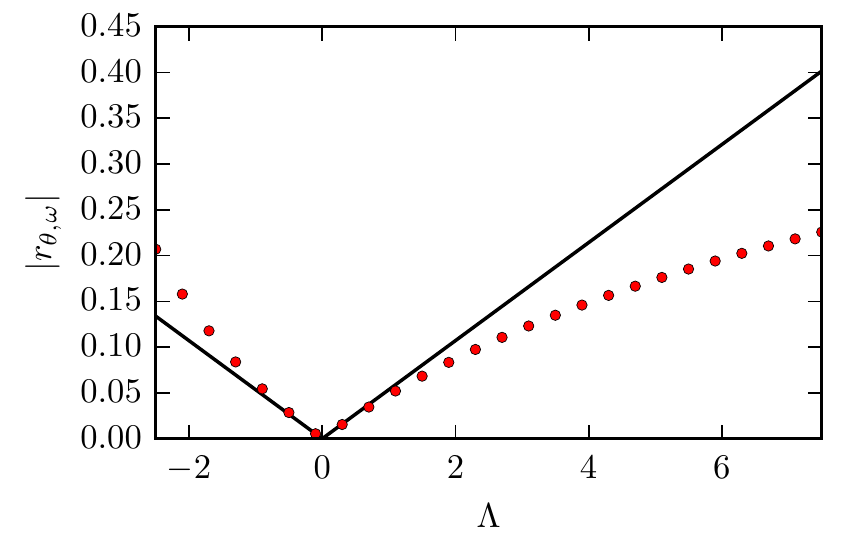}\put(2,150){(a)}\end{overpic}
\begin{overpic}[width=\linewidth]{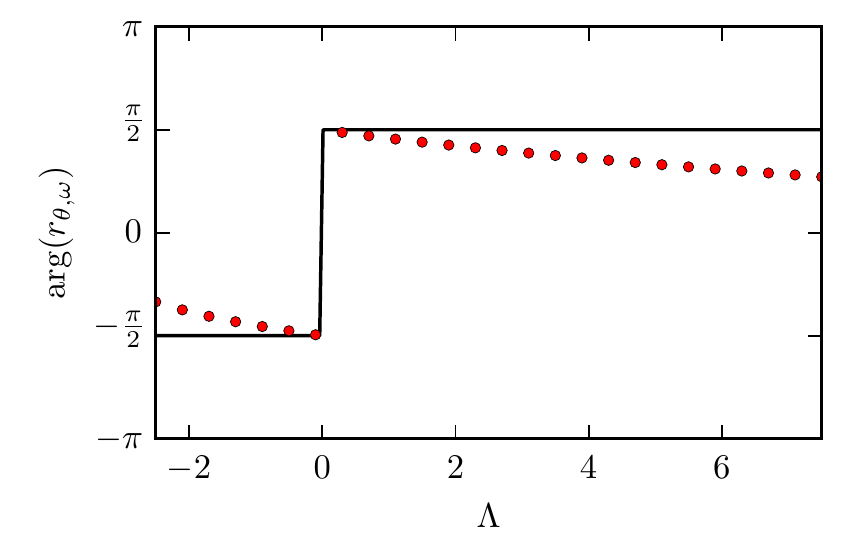}\put(2,150){(b)}\end{overpic}
\begin{overpic}[width=\linewidth]{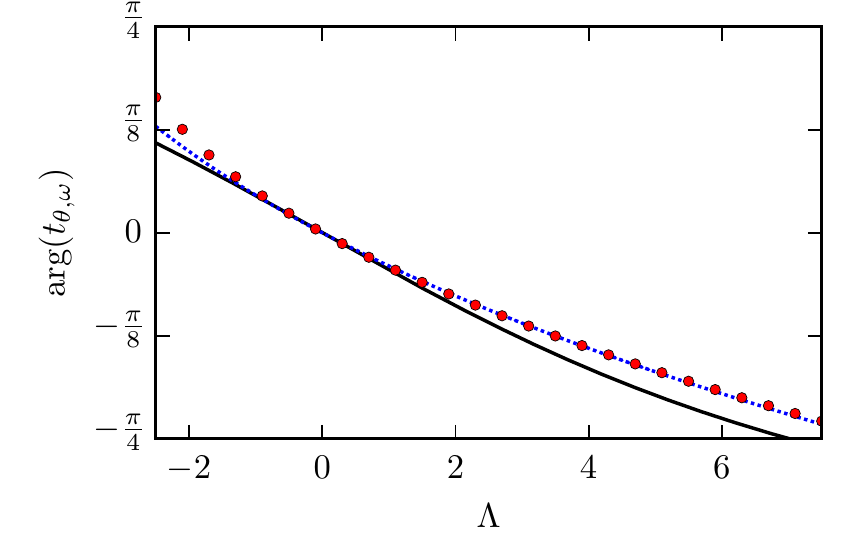}\put(2,150){(c)}\end{overpic}
\caption{\label{fig:vslambda} (Color online) Coupling constant dependence of $|r|$, $\arg(r)$, and $\arg(t)$. Panel (a) The quantity $|r|$ as a function of $\Lambda$, for $q_{\rm pl}d=0.23$. For this value of $q_{\rm pl}d$, the reflection coefficient displays a maximum. Red circles are numerical data (including nonlocal corrections) while the solid black line is the result of the first-order Born approximation. Panel (b) Same as in panel (a) but for the phase of the reflection coefficient. Panel (c) Same as in panel (b) for $\arg(t)$. In this panel, the blue dashed line is the result of the eikonal approximation.}
\end{figure}

In Fig.~\ref{fig:vslambda} we illustrate the dependence of $|r|$, $\arg(r)$, and $\arg(t)$ on the coupling constant $\Lambda$. Full numerical results (symbols) are compared with the results of the first-order Born and eikonal approximations. The perturbative validity of the former is again clear. The validity of the eikonal approximation for $\arg(t)$ and its non-perturbative nature are also clear. 

\begin{figure}
\begin{overpic}[width=\linewidth]{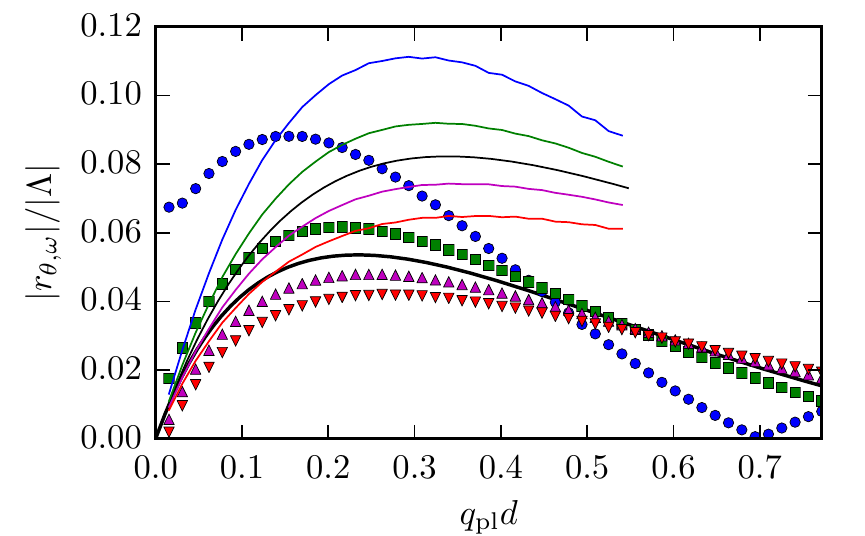}\put(2,150){(a)}\end{overpic}
\begin{overpic}[width=\linewidth]{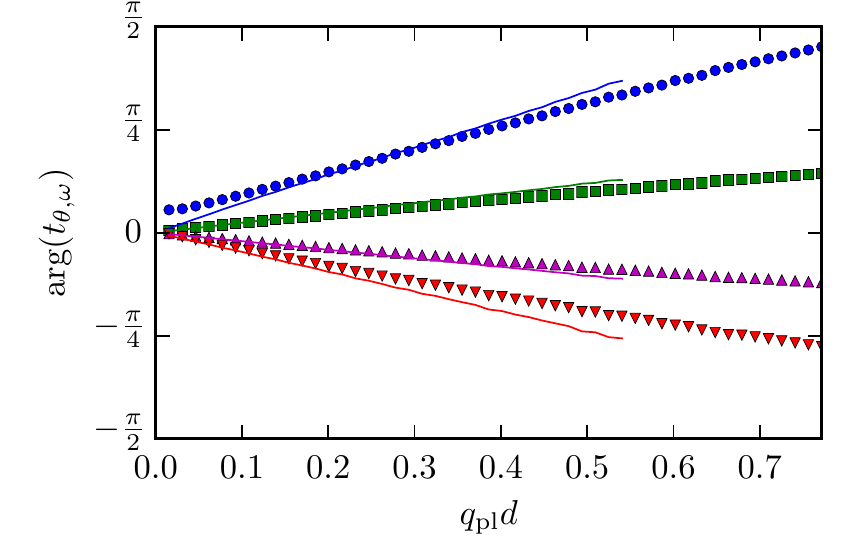}\put(2,150){(b)}\end{overpic}
\caption{\label{fig:localtheory} (Color online) Numerical results of the full nonlocal theory (symbols) are compared with the corresponding results of the local theory (thin lines). Color coding is identical to that used in Fig.~\ref{fig:amplitudes}. The black line in panel (a) is the result of the first-order Born approximation (which is ``universal'', provided that $|r_{\theta,\omega}|$ is rescaled by $|\Lambda|$). As expected, the local theory is a good approximation in the long-wavelength $q_{\rm pl}d \to 0$ limit.}
\end{figure}

Finally, in Fig.~\ref{fig:localtheory} we compare our full numerical results with the results of the local theory, which is obtained by setting $C(\omega)\equiv 2\pi e^2/\bar{\epsilon}$, ${\cal W}(q,\omega)\equiv 0$, and ${\cal K}(q_x,q_x',q_y,\omega)\equiv 1$ in the general equations. As expected, the local theory fails spectacularly in predicting $|r|$ for large values of the product $q_{\rm pl} d$.

\section{Summary and conclusions}
\label{sect:conclusions}

In summary, we have presented a general theoretical framework to calculate the scattering properties of 2D plasmons against perturbations coupling to density, current, and real-spin operators. The theory discussed in this Article differs from other theories based on Maxwell equations combined with local/phenomenological approximations for the spatial dependence of the conductivity: i) it is essentially semi-analytical, requires little numerical effort, and takes into account nonlocal effects; ii) instead of assuming a phenomenological model for the spatial dependence of the conductivity profile, it relies on microscopic calculations of the density-density response function for a given Hamiltonian in the presence of external fields; iii) finally, it treats on equal footing a wide variety of perturbations. 

We have discussed in great detail the case of parabolic-band 2D electron systems, such as those that can be found in high-mobility heterostructures based on GaAs. Due to subtleties of the massless Dirac fermion Hamiltonian (i.e.~presence of inter-band particle-hole excitations of arbitrarily large energy), the case of plasmon scattering in a doped graphene sheet has been discussed only for external perturbations that couple to the density operator and is presented in Appendix~\ref{sect:graphene}.

Our theory starts from a Lippmann-Schwinger equation for the screened potential 
$V_{\rm sc}(\bm q, \omega)$ in an inhomogeneous 2D electron system---see Eq.~(\ref{eq:lippmannschwinger}). The key unknown quantity in this equation is the scattering kernel, $\Delta(\bm q, \bm q', \omega)$, which is defined in terms of the density-density response function of the inhomogeneous 2D electron system in Eq.~(\ref{eq:Udefinition}). The latter input is calculated analytically in the long-wavelength limit, the key information being encoded in the so-called moments $M^{(1)}(\bm q, \bm q')$, $M^{(2)}(\bm q, \bm q')$, and $M^{(3)}(\bm q, \bm q')$, which are explicitly reported in Sect.~\ref{sect:chi}. For the reasons stated above, the case of a doped graphene sheet is separately discussed in Sect.~\ref{sect:graphene}. Crucially, the density-density response function is calculated transcending the usual local approximation.

In Sect.~\ref{sect:example} we have reported illustrative numerical results for the scattering of 2D plasmons against a single point-like charged impurity and a 1D electrostatic barrier due to a line of charges. The solutions of these two problems are mainly used to highlight i) the range of validity of the Born and eikonal approximations (with respect to the exact numerical solution of the Lippmann-Schwinger equation) and ii) to stress the importance of nonlocal effects. We emphasize, for the sake of completeness, that the present theory has also been very successfully used to explain experimental data related to a plasmonic phase shifter realized by using encapsulated graphene~\cite{woessner}. 

In the future, we plan to discuss examples in which a dielectric perturbation causing a change $\delta v(\bm q,\bm q',\omega)$ in the Coulomb interaction is present, 
to extend the graphene theory of Sect.~\ref{sect:graphene} to a larger variety of perturbations, and to deal with the case of intense perturbations.
\acknowledgments
We acknowledge Achim Woessner and Frank Koppens for many useful and inspiring discussions. I.T. and M.P. wish to thank Michael Beconcini for early contributions to this work and Andrea Tomadin for useful discussions. This work was supported by the European Union's Horizon 2020 research and innovation programme under grant agreement No.~696656 ``GrapheneCore1'', Fondazione Istituto Italiano di Tecnologia,   and the ERC Advanced Grant 338957 FEMTO/NANO. M.P. is extremely grateful for the financial support granted by ICFO during a visit in August 2016.
\appendix
\section{Non-interacting density-density response function of an inhomogeneous electron liquid}
\label{sect:chi}
The proper density-density response function of the inhomogeneous 2D electron system under study is a crucial input for the microscopic calculation of the scattering kernel in Eq.~(\ref{eq:Udefinition}).

In this Section we first consider a parabolic-band 2D electron gas~\cite{Giuliani_and_Vignale} subject to {\it three} different types of perturbations that break translational and rotational invariance. We calculate the proper density-density response function within the aforementioned RPA~\cite{Giuliani_and_Vignale}.
Our results have the form of a rigorous expansion in inverse powers of the frequency $\omega$. We calculate the leading and next-to-leading terms of this expansion. Section~\ref{sect:graphene} below will be devoted to the case of an inhomogeneous graphene sheet.

The proper density-density response function $\tilde{\chi}_{nn}(\bm q,\bm q',\omega)$ is defined in Eq.~(\ref{eq:induceddensity}). In the RPA, this complicated function is brutally replaced by the density-density response function $\chi_{\rm H}(\bm q,\bm q',\omega)$ of a formally non-interacting system usually termed the ``Hartree system''~\cite{Giuliani_and_Vignale}. In the case of a 2D parabolic-band electron gas, the energy eigenstates of the Hartree system are determined by the Hamiltonian
\begin{equation}\label{eq:hamiltonian}
{\cal H}=\frac{1}{2 m}\left[{\bm p}+\frac{e}{c} {\bm A}({\bm r}) \right]^2+U_{\rm tot}({\bm r})+{\bm Z}({\bm r})\cdot {\bm \sigma}~.
\end{equation}
Here $m$ is the electron band mass, $-e$ the electron charge, $c$ the speed of light, 
\begin{equation}
U_{\rm tot}({\bm r}) = U_{\rm ext}({\bm r}) + U_{\rm H}({\bm r})
\end{equation}
is the sum of an external scalar potential $U_{\rm ext}({\bm r})$ and the self-consistent Hartree potential~\cite{Giuliani_and_Vignale} $U_{\rm H}({\bm r})$, 
${\bm A}({\bm r})$ is an external vector potential, and, finally, ${\bm Z}({\bm r})$ is an external Zeeman field that couples to spin degrees of freedom. In Eq.~(\ref{eq:hamiltonian}), ${\bm \sigma}$ is a vector of spin-$1/2$ Pauli matrices, $\sigma^\alpha$ with $\alpha=x,y,z$. Without loss of generality, we work in the Coulomb gauge for the vector potential, i.e.~$\nabla \cdot \bm A(\bm r)=0$. In 2D, the electron orbital motion is influenced only by the perpendicular component of the magnetic field $B_z({\bm r})$, while the Zeeman field ${\bm Z}({\bm r})$ couples to all three components of the electron's spin.

For homogeneous electron systems $U_{\rm H}$ is cancelled exactly by the background potential and the Hartree system reduces to the corresponding homogeneous non-interacting electron system~\cite{Giuliani_and_Vignale}.

Following Ref.~\onlinecite{Giuliani_and_Vignale}, we express $\chi_{\rm H}(\bm q,\bm q',\omega)$ in terms of a density-density correlator at different times:
\begin{equation}\label{eq:kubo}
\chi_{\rm H}(\bm q, \bm q',\omega)=\frac{-i}{S\hbar}\lim_{\eta \to 0^+}\int_0^\infty d \tau \,e^{i(\omega+i\eta)\tau}\langle [n_{\bm q}(\tau), n_{-\bm q'}]\rangle~,
\end{equation}
where the density operator at time $\tau$ is defined via the usual Heisenberg time evolution operator, i.e.~$n_{\bm q}(\tau) = \exp(i {\cal H} \tau/\hbar)n_{\bm q} \exp(-i {\cal H} \tau/\hbar)$, and 
$n_{\bm q}$ is given by
\begin{equation}
n_{\bm q}=e^{-i \bm q \cdot {\bm r}}~.
\end{equation}

In Eq.~(\ref{eq:kubo}), the average must be taken over the ground state of the Hamiltonian~(\ref{eq:hamiltonian}) and $[\dots, \dots]$ denotes a commutator. Now, the key point is that this average can be expanded~\cite{Giuliani_and_Vignale} in a Taylor series for small values of $\tau$:
\begin{equation}\label{eq:short-time-expansion}
\langle [n_{\bm q}(\tau), n_{-\bm q'}]\rangle=\sum_{\ell=0}^{\infty}\frac{\tau^{\ell}}{\ell!}
\langle [n_{\bm q}^{(\ell)} ,n_{-\bm q'}]\rangle~,
\end{equation}
where, for any operator ${\cal O}$,
\begin{equation}\label{eq:zeroth-order}
{\cal O}^{(0)} \equiv \cal O~,
\end{equation}
and for integer values of $\ell \geq 1$ the $\ell$-th time derivative of the operator ${\cal O}$ is defined by
\begin{equation}
{\cal O}^{(\ell)} \equiv \frac{i}{\hbar}[{\cal H}, {\cal O}^{(\ell - 1)}]~.
\end{equation}

In the following, we make use of three useful identities involving time derivatives of operators. 
For any $\ell \geq 1$ the expectation value of the $\ell$-th time derivative of an operator, calculated over the ground state (or any other equilibrium state), vanishes:
\begin{equation}\label{eq:derivativeatequilibrium}
\langle {\cal O}^{(\ell)} \rangle=\frac{i}{\hbar}\langle{\cal H} {\cal O}^{(\ell - 1)}-{\cal O}^{(\ell - 1)}{\cal H}\rangle=0~.
\end{equation}
Products of operators are differentiated with respect to time according to the Leibniz rule:
\begin{equation}\label{eq:derivationproduct}
\begin{split}
\left( A B\right)^{(1)} & =\frac{i}{\hbar}\left({\cal H}AB-A{\cal H}B+A{\cal H}B-AB{\cal H}\right)\\
 & =A^{(1)}B^{(0)}+A^{(0)}B^{(1)}~.
\end{split}
\end{equation}
Combining Eq.~(\ref{eq:derivativeatequilibrium}) and~(\ref{eq:derivationproduct}), we obtain the ``integration by parts'' rule:
\begin{equation}\label{eq:integrationbyparts}
\langle A^{(n+1)}B^{(m)}\rangle=-\langle A^{(n)}B^{(m+1)} \rangle~.
\end{equation}

After integration over time $\tau$, Eq.~(\ref{eq:short-time-expansion}) translates into an expansion of $\chi_{\rm H}(\bm q,\bm q',\omega)$ in inverse powers of $\omega$:
\begin{equation}\label{eq:powerexpansion}
\chi_{\rm H}(\bm q, \bm q',\omega) = \sum_{\ell=0}^{\infty}\frac{M^{(\ell)}(\bm q,\bm q')}{\omega^{\ell+1}}
\end{equation}
where
\begin{equation}\label{eq:moments}
M^{(\ell)}(\bm q,\bm q')\equiv\frac{i^{\ell}}{\hbar S} \langle [n_{\bm q}^{(\ell)} ,n_{-\bm q'}] \rangle~.
\end{equation}
The following reciprocity relations hold for the coefficients $M^{(\ell)}(\bm q, \bm q')$:
\begin{equation}
M^{(2\ell)}(\bm q,\bm q')=- [M^{(2\ell)}(-\bm q,-\bm q')]^{*}~,
\end{equation}
\begin{equation}
M^{(2\ell + 1)}(\bm q,\bm q') = [M^{(2\ell + 1)}(-\bm q,-\bm q')]^{*}~,
\end{equation}
and
\begin{equation}\label{eq:time-inversion}
M^{(\ell)}(\bm q,\bm q')=[M^{(\ell)}(-\bm q',-\bm q)]_{t}~.
\end{equation}
In Eq.~(\ref{eq:time-inversion}), $[\dots]_t$ represents time inversion.

We now proceed to calculate the coefficients $M^{(\ell)}(\bm q,\bm q')$ of the expansion in Eq.~(\ref{eq:powerexpansion}) up to $\ell=3$ for the system described by the Hamiltonian~(\ref{eq:hamiltonian}).

For later convenience, we introduce the {\it kinetic} momentum operator ${\bm \Pi}$ with Cartesian components
\begin{equation}
\Pi_\alpha \equiv p_\alpha+\frac{e}{c} A_{\alpha}({\bm r})~.
\end{equation}
The kinetic momentum operator has the same commutation relation with the position operator as the {\it canonical} momentum operator,
\begin{equation}\label{eq:comm1}
[r_\alpha, \Pi_\beta]=i\hbar \delta_{\alpha \beta}~.
\end{equation}
However, different Cartesian components of ${\bm \Pi}$ do not commute with each other~\cite{Giuliani_and_Vignale}:
\begin{equation}\label{eq:comm2}
[\Pi_\alpha,\Pi_\beta]=-\frac{ie\hbar}{c}\epsilon_{\alpha \beta}\partial_\alpha A_\beta({\bm r})=-\frac{ie\hbar}{c}B_z({\bm r})~,
\end{equation}
where $B_z({\bm r})$ is the magnetic field in the $\hat{\bm z}$ direction at position $\bm r$ in space. A sum over repeated indices is intended in Eq.~(\ref{eq:comm2}) and below.

Introducing the kinetic momentum operator, we can rewrite the Hamiltonian (\ref{eq:hamiltonian}) in the following manner:
\begin{equation}\label{eq:hamiltonian2}
{\cal H}=\frac{1}{2 m}\Pi_\alpha \Pi_\alpha + U_{\rm tot}({\bm r})+Z_{\alpha}({\bm r})\otimes \sigma^{\alpha}~.
\end{equation}
\subsection{Calculation of $M^{(0)}$}
As we have seen above in Eq.~(\ref{eq:zeroth-order}), the zeroeth-order derivative of an operator coincides with the operator itself: $n_{\bm q}^{(0)}= n_{\bm q}$.
Density operators commute among each other because they are functions of the position operator only,
\begin{equation}\label{eq:comm3}
[n_{\bm q}, n_{-\bm q'}]=0~.
\end{equation}
We therefore conclude that $M^{(0)}(\bm q,\bm q')$ vanishes identically.

\subsection{Calculation of $M^{(1)}$}

The first non-trivial term of the expansion (\ref{eq:powerexpansion}) is determined by $M^{(1)}$.
The time derivative of the density operator can be easily calculated:
\begin{equation}\label{eq:continuity}
n_{\bm q}^{(1)}=\frac{i}{\hbar} \left[{\cal H}, n_{\bm q}\right]=\frac{i}{2\hbar m }\left[\Pi_{\alpha} \Pi_{\alpha}, n_{\bm q}\right]=-iq_\alpha J_{\bm q, \alpha}~.
\end{equation}
In deriving the second equality we made use of the fact that the scalar and Zeeman terms of the Hamiltonian (\ref{eq:hamiltonian2}) 
commute with the density operator. In deriving the third equality, we made use of the following commutator
\begin{equation}\label{eq:comm4}
[\Pi_\alpha, n_{\bm q} ]= [p_\alpha, n_{\bm q}]=-\hbar q_\alpha n_{\bm q}
\end{equation}
and introduced the {\it physical} (i.e.~gauge-invariant) current operator in Fourier space:
\begin{eqnarray}\label{eq:currentdef}
J_{\bm q, \alpha}&=&\frac{1}{2m}\lbrace \Pi_\alpha, n_{\bm q}\rbrace~,
\end{eqnarray}
where $\{\dots,\dots\}$ denotes an anticommutator. Note that Eq.~(\ref{eq:continuity}) is the operator version of the continuity equation.

Using Eqs.~(\ref{eq:comm4})-(\ref{eq:currentdef}) we calculate $M^{(1)}(\bm q,\bm q')$ finding:
\begin{equation}\label{eq:M1}
M^{(1)}(\bm q,\bm q')=\frac{i}{\hbar S}\langle [-iq_\alpha J_{\bm q, \alpha}, n_{-\bm q'}]\rangle=\frac{q_\alpha q_\alpha'}{mS} \langle n_{\bm q-\bm q'}\rangle~.
\end{equation}
For $\bm q=\bm q'$, Eq.~(\ref{eq:powerexpansion}) for $\ell=1$ and~Eq.~(\ref{eq:M1}) reduce to the usual f-sum rule for homogeneous electron systems~\cite{Pines_and_Nozieres,Giuliani_and_Vignale}.
\subsection{Calculation of $M^{(2)}$}
Using Eq.~(\ref{eq:continuity}), we immediately see that the second derivative of the density operator is proportional to the first derivative of the current operator, 
\begin{equation}\label{eq:n2}
n^{(2)}_{\bm q} =-iq_\alpha J_{\bm q, \alpha}^{(1)}~.
\end{equation}
The latter can be calculated by repeated use of the commutation relations (\ref{eq:comm1}), (\ref{eq:comm2}), (\ref{eq:comm3}), and~(\ref{eq:comm4}) and reads as following:
\begin{equation}\label{eq:NavierStokes}
\begin{split}
J_{\bm q, \alpha}^{(1)}= &  -\frac{i q_\beta}{m}\left(T_{\bm q, \alpha \beta}-\frac{\hbar^2 q_\alpha q_\beta}{4m} n_{\bm q}\right)\\
& +\frac{1}{m} F_{\bm q, \alpha}-\frac{1}{m}\partial_\alpha U_{\rm tot} n_{\bm q}-\frac{1}{m}\partial_\alpha Z_\beta S_{\bm q, \beta}~.
\end{split}
\end{equation}
In Eq.~(\ref{eq:NavierStokes}) we introduced the stress tensor operator
\begin{equation}\label{eq:stresstensordef}
T_{\bm q, \alpha \beta} \equiv \frac{1}{4m} \lbrace \lbrace \Pi_\alpha, \Pi_\beta\rbrace, 
n_{\bm q}\rbrace~,
\end{equation}
the Lorentz force-density operator
\begin{equation}
F_{\bm q, \alpha} \equiv -\frac{e \epsilon_{\alpha \beta}}{4 mc} \lbrace \lbrace \Pi_\beta , B_z(\bm r)\rbrace, n_{\bm q}\rbrace~,
\end{equation}
and the spin-density operator
\begin{equation}
S_{\bm q, \alpha} \equiv n_{\bm q}\sigma^{\alpha}~.
\end{equation}

Taking the commutator in Eq.~(\ref{eq:moments}) with the density operator at wavevector $-\bm q'$ we obtain the final result
\begin{eqnarray}\label{eq:M2}
M^{(2)}(\bm q,\bm q') & = & q_\alpha q_\alpha' q_\beta \frac{2}{mS}\langle J_{\bm q-\bm q', \beta}
\rangle \nonumber\\
& - & \epsilon_{\alpha \beta}q_\alpha q_\beta' \frac{ie}{m^2cS} \langle B_z({\bm r}) n_{\bm q-\bm q'}\rangle\nonumber\\
& = & \frac{q_\alpha q_\alpha'(q_\beta +q_\beta')}{mS} \langle J_{\bm q-\bm q', \beta}\rangle \nonumber\\
& - & \epsilon_{\alpha \beta}q_\alpha q_\beta' \frac{ie}{m^2cS} \left\langle B_z({\bm r}) n_{\bm q-\bm q'}\right\rangle~.
\end{eqnarray}
In the second step we used the continuity equations (\ref{eq:continuity}) and the identity (\ref{eq:derivativeatequilibrium}) (the aim of this manipulation was to put the result in a more symmetric form). 
\subsection{Calculation of $M^{(3)}$}
The calculation of the third moment, $M^{(3)}(\bm q,\bm q')$, is quite cumbersome. It can be simplified by using an ``integration by parts'' described in Eq.~(\ref{eq:integrationbyparts}), together with Eqs.~(\ref{eq:n2}) and~(\ref{eq:continuity}):
\begin{equation}
\begin{split}
M^{(3)}(\bm q,\bm q') & = 
-\frac{i}{\hbar S} \langle [n^{(3)}_{\bm q}, n_{-\bm q'}]\rangle
=\frac{i}{\hbar S} \langle [n^{(2)}_{\bm q}, n^{(1)}_{-\bm q'}]\rangle\\
& =\frac{i}{\hbar S}q_\alpha q_\beta' \langle [J^{(1)}_{\bm q, \alpha}, J_{-\bm q', \beta}]\rangle~.
\end{split}
\end{equation}
We then evaluate the commutator $[J^{(1)}_{\bm q, \alpha}, J_{-\bm q', \beta}]$ by using Eqs.~(\ref{eq:currentdef})-(\ref{eq:NavierStokes}) and the commutation rules (\ref{eq:comm1}), (\ref{eq:comm2}), (\ref{eq:comm3}), and~(\ref{eq:comm4}). We find
\begin{eqnarray}\label{eq:M3}
M^{(3)}(\bm q,\bm q') & = & q_\gamma q_\gamma'q_\alpha q_\beta' \frac{3}{m^2S} 
\langle T_{\bm q-\bm q', \alpha \beta}\rangle + q_\alpha q_\alpha' \frac{\hbar^2}{m^3 S} \nonumber\\
&\times &\Big\{\frac{3}{4}[q_\beta(q_\beta-q_\beta')][q_\beta'(q_\beta-q_\beta')] +\frac{1}{4}(q_\beta q_\beta')^2\Big\} \nonumber\\
&\times& \langle n_{\bm q-\bm q'}\rangle \nonumber\\
&+ & q_\alpha q_\beta'\frac{1}{m^2S}\langle \partial_\alpha \partial_\beta U_{\rm tot}({\bm r}) n_{\bm q-\bm q'}
\rangle \nonumber\\
& + & q_\alpha q_\beta' \frac{1}{m^2S} \langle \partial_\alpha \partial_\beta Z_\gamma({\bm r}) 
S_{\bm q-\bm q', \gamma} \rangle\nonumber \\
&+ & q_\alpha q_\alpha'\frac{e^2}{m^3c^2S} \langle B^2_z({\bm r}) n_{\bm q-\bm q'}\rangle\nonumber\\
& + & \frac{3}{2}\left(q_\beta q_\beta' q_\alpha - q^2 q'_\alpha\right)\frac{i}{m^2 S}
\langle F_{\bm q-\bm q', \alpha}\rangle \nonumber\\
& + & q_\alpha q_\gamma' \epsilon_{\alpha \beta}\frac{e}{m^2 c S}
\langle L_{\bm q-\bm q', \beta \gamma}\rangle \nonumber \\
& +& q_\alpha q_\beta ' \epsilon_{\alpha \beta}\frac{e}{m^2 cS}
\langle L_{\bm q-\bm q', \gamma \gamma}\rangle~.
\end{eqnarray}
Because of the presence of a non-uniform magnetic field, Eq.~(\ref{eq:M3}) contains two terms that involve 
the tensor
\begin{equation}
L_{\bm q, \alpha \beta}=\frac{1}{4m}\lbrace\lbrace \Pi_\alpha,\partial_\beta B_z({\bm r})\rbrace , n_{\bm q}\rbrace~.
\end{equation}
\section{On inhomogeneous 2D electron systems in graphene}
\label{sect:graphene}

The technique used in the previous Section to calculate the density-density response function of an inhomogeneous parabolic-band electron liquid cannot be applied directly to graphene.
The main reason is that the high-frequency expansion (\ref{eq:powerexpansion}) is invalidated by the presence of particle-hole excitations of arbitrarily large energy, which are associated to inter-band transitions~\cite{abedinpour_prb_2011,sabio_prb_2008}.

In this case, instead of calculating microscopically the density-density response function of an inhomogeneous system of 2D massless Dirac fermions~\cite{kotov_rmp_2012}, we choose a more humble approach. We find a semi-phenomenological expression for $\chi_{\rm H}(\bm q,\bm q',\omega)$ which is able to capture, even if in an approximate way, nonlocal effects. The functional dependence of $\chi_{\rm H}(\bm q,\bm q',\omega)$ on wavevectors and frequency is chosen is such a way that it respects the following requirements: 
(i) it is equivalent to a local spatially-dependent conductivity~\cite{MartinMorenoACSnano2013} at second order in the wavevectors; (ii) it reduces to 
\begin{equation}
\chi_{\rm H}(\bm q,\bm q',\omega) =\delta_{\bm q, \bm q'}\left[\frac{E_{\rm F}q^2}{\pi  \hbar^2 \omega^2}
-\frac{E_{\rm F}^{-1}q^2}{4 \pi}+\frac{3 v_{\rm F}^2 E_{\rm F}q^4}{4\pi \hbar^2 \omega^4}\right]
\end{equation}
in the homogeneous limit; (iii) it contains terms ${\cal O}(q^4)$ (i.e.~it takes into account nonlocal effects); 
(iv) it depends only on the spatially-dependent ground-state electron density $n({\bm r})$.

We propose the following simple expression that meets all the requirements (i)-(iv):
\begin{equation}\label{eq:chinonhomogeneous}
\begin{split}
\chi_{\rm H}(\bm q,\bm q',\omega) & =\frac{E_{\rm F}(\bm q-\bm q')\bm q \cdot \bm q'}{\pi S \hbar^2 \omega^2}
-\frac{E_{\rm F}^{-1}(\bm q-\bm q')\bm q \cdot \bm q'}{4 \pi S}\\
& +\frac{3 v_{\rm F}^2 E_{\rm F}(\bm q-\bm q')(\bm q \cdot \bm q')^2}{4\pi S\hbar^2 \omega^4}~.
\end{split}
\end{equation}
Here $E_{\rm F}(\bm q)$ is defined by
\begin{equation}
E_{\rm F}(\bm q)\equiv\int d \bm r~e^{-i \bm q \cdot \bm r} \hbar v_{\rm F}~{\rm sgn}[n(\bm r)]\sqrt{\pi |n(\bm r)|}~,
\end{equation}
where $v_{\rm F}$ is the graphene Fermi velocity~\cite{kotov_rmp_2012} and $n(\bm r)$ the local carrier density. Similarly,
\begin{equation}
E_{\rm F}^{-1}(\bm q)\equiv \int d \bm r~e^{-i \bm q \cdot \bm r} \left[\hbar v_{\rm F}~{\rm sgn}[n(\bm r)]\sqrt{\pi |n(\bm r)|}\right]^{-1}~.
\end{equation}

Eq.~(\ref{eq:chinonhomogeneous}) has been successfully used in Ref.~\onlinecite{woessner} to calculate the the transmission coefficient in an experimental geometry of the type sketched in Fig.~\ref{fig:geometry}(b).
\section{Effective interaction}
\label{app:effectiveinteraction}
The effective interaction satisfies Eq.~(\ref{eq:effectiveinteraction}).
Assuming that $W_{\rm h}^{-1}(q,\omega)$ has only one simple zero at $q_{\rm pl}$, the most general solution of this {\it distributional} equation is
\begin{equation}\label{eq:solutioneffectiveinteraction} 
W_{\rm h}(q,\omega)=
\mathcal{P} \frac{1}{W_{\rm h}^{-1}(q,\omega)}+A \delta(q-q_{\rm pl})~.
\end{equation}
Here $\mathcal{P}$ stands for Cauchy principal value, while the constant $A$ must be chosen to satisfy the required boundary conditions. To see what is the correct choice of $A$ to have an outgoing wave, we look at the asymptotic behavior of $W_{\rm h}$ in real space for large $r$.

To begin with, we can single out the divergent part of the interaction 
by rewriting Eq.~(\ref{eq:solutioneffectiveinteraction}) as 
\begin{equation}\label{eq:solutioneffectiveinteraction2}
\begin{split}
&W_{\rm h}(q,\omega)=\\
&C(\omega)\left[\mathcal{P}\frac{1}{q_{\rm pl}-q}+B \delta(q_{\rm pl}-q)+\frac{1}{q}+\frac{\mathcal{W}(q,\omega)}{q_{\rm pl}}\right]~.
\end{split}
\end{equation}
Since the last term in Eq.~(\ref{eq:solutioneffectiveinteraction2}) is regular at $q=q_{\rm pl}$, it does not affect the asymptotic behavior of the effective interaction at large distances, i.e.~for $r q_{\rm pl}(\omega)\gg1$. 
We can therefore write
\begin{equation}\label{eq:solutioneffectiveinteraction3} 
\begin{split}
& W_{\rm h}(r \gg q_{\rm pl}^{-1},\omega) \\
& \simeq C(\omega)\int \frac{d \bm q}{(2\pi)^2}e^{i \bm q \cdot \bm r}\left[\mathcal{P}\frac{1}{q_{\rm pl}-q}+B \delta(q_{\rm pl}-q)+\frac{1}{q}\right]\\
& =C(\omega)q_{\rm pl}\left\lbrace\frac{1}{2}\y0(q_{\rm pl}r)+\frac{B}{2\pi} \j0(q_{\rm pl}r)\right. \\
& \left. +\frac{1}{4}\left[\H0( q_{\rm pl}r)-\y0(q_{\rm pl}r)\right]\right\rbrace~,
\end{split}
\end{equation}
where $\j0(x)$ and $\y0(x)$ are the first and second kind Bessel functions and $\H0(x)$ is the Struve function.
The term in square bracket goes to zero like $r^{-1}$ for large $r$, while $\j0(x)$ and $\y0(x)$ are oscillating functions whose amplitudes decay like $r^{-1/2}$ for large $r$.

The correct combination for an outgoing (ingoing) cylindrical wave is obtained by setting $B=\mp \pi i$, a choice which yields the first (second) Hankel function $\h012(x)=\j0(x) \pm i \y0(x)$. With this choice of $B$ the asymptotic behavior of $W_{\rm h}$ is
\begin{equation}\label{eq:effectiveinteractionbehaviour} 
\begin{split}
W_{\rm h}^{(\pm)}(r\gg q_{\rm pl}^{-1},\omega) & \simeq \frac{\mp i}{2}q_{\rm pl}C(\omega)\h012 (q_{\rm pl}r)\\
& \simeq \frac{\mp i}{\sqrt{2\pi r}}\sqrt{q_{\rm pl}}C(\omega)e^{\pm i(q_{\rm pl}r-\frac{\pi}{4})}~,
\end{split}
\end{equation}
where we dropped all terms decaying faster than $r^{-1/2}$ and made use of the asymptotic behavior of Hankel's functions.

If we neglect the correction ${\cal W}$, the effective interaction in real space is
\begin{equation}\label{eq:Wapproxrealspace} 
\begin{split}
& W_{\rm h}^{(\pm)}(r,\omega) = \\
&C(\omega)q_{\rm pl}\left\lbrace \mp \frac{i}{2}\h012( q_{\rm pl}r)+\frac{1}{4}\left[\H0(q_{\rm pl}r)-\y0(q_{\rm pl}r)\right] \right\rbrace~.
\end{split}
\end{equation}
\section{Real space formulation of the scattering equations}
\label{app:realspace}
\subsection{Geometry in Fig.~\ref{fig:geometry}(a)}
To make a connection between the scattering amplitude $f(\theta_{\bm r},\theta,\omega)$ and the solutions of the scattering equation in momentum space it is useful to rewrite the latter equation in real space.
The real-space version of Eq.~(\ref{eq:scatteringequation}) reads as following:
\begin{equation}\label{eq:scatteringrealspace}
\begin{split}
&\int d^2 \bm r' W_{\rm h}^{-1}(|\bm r -\bm r'|,\omega) V_{\rm sc}(\bm r',\omega)=\\
& \frac{1}{S}\int d^2 \bm r' \Delta(\bm r,\bm r',\omega)V_{\rm sc}(\bm r',\omega)~,
\end{split}
\end{equation}
while Eq.~(\ref{eq:lippmannschwinger}) becomes
\begin{equation}\label{eq:lipmannschwingerrealspace}
\begin{split}
V_{\rm sc}(\bm r,\omega)=V^{(0)}(\bm r,\omega)+\int d \bm r'W_{\rm h}^{(+)}(|\bm r-\bm r'|,\omega)\\
\times  \frac{1}{S} \int d \bm r'' \Delta(\bm r',\bm r'',\omega)V_{\rm sc}(\bm r'',\omega)~.
\end{split}
\end{equation}
Since the inhomogeneity is localized in a finite region of space of radius $a$, $\Delta(\bm r,\bm r',\omega)$ vanishes if $r$ or $r'$ are bigger than $a$.
For $r\gg a, q_{\rm pl}^{-1}$ we can also approximate $|\bm r-\bm r'|\simeq r-\bm r \cdot \bm r'/r$. 
Setting $V^{(0)}(\bm r,\omega)=\exp{(i \bm q_{\rm pl}\cdot \bm r)}$ we obtain
\begin{eqnarray}\label{eq:lipmannscwingerasymptotic}
 V_{\rm sc}(\bm r,\omega) &\simeq & e^{i \bm q_{\rm pl} \cdot \bm r} + \int_{r'<a} d^{2}\bm r'W_{\rm h}^{(+)}\left(r-\hat{\bm r} \cdot \bm r',\omega\right) \nonumber\\
 &\times&\int_{r''<a} d^{2} \bm r'' \frac{1}{S}\Delta(\bm r',\bm r'',\omega)V_{\rm sc}(\bm r'',\omega) 
 \nonumber\\
& \simeq & e^{i \bm q_{\rm pl} \cdot \bm r} - C(\omega)\int_{r'<a} d^{2}\bm r'\frac{i\sqrt{q_{\rm pl}}}{\sqrt{2\pi r}}e^{i[q_{\rm pl}(r-\hat{\bm r}\cdot \bm r')-\frac{\pi}{4}]} \nonumber\\
&\times&\int_{r''<a} d^{2} \bm r'' \frac{1}{S} \Delta(\bm r',\bm r'',\omega)V_{\rm sc}(\bm r'',\omega)\nonumber\\
& \simeq& e^{i \bm q_{\rm pl}\cdot \bm r} - \frac{\sqrt{q_{\rm pl}} e^{i\frac{\pi}4}}{\sqrt{2\pi r}}C(\omega)e^{iq_{\rm pl}r}
\int_{r'<a} d^{2}\bm r' e^{-iq_{\rm pl}\hat{\bm r}\cdot \bm r'}
\nonumber\\
&\times&\int_{r''<a} d^{2} \bm r'' \frac{1}{S} \Delta(\bm r',\bm r'',\omega)V_{\rm sc}(\bm r'',\omega)
\nonumber\\
& \simeq & e^{i \bm q_{\rm pl}\cdot \bm r} -\frac{\sqrt{q_{\rm pl}}e^{i\frac{\pi}4}}{\sqrt{2\pi r}}C(\omega)e^{iq_{\rm pl}r}\nonumber\\
&\times&\int_{r'<a} d^{2} \bm r' e^{-iq_{\rm pl}\hat{\bm r}\cdot \bm r'}T(\bm r',\theta,\omega) \nonumber\\
&\simeq& e^{i \bm q_{\rm pl}\cdot \bm r} -\frac{\sqrt{q_{\rm pl}}e^{i\frac{\pi}4}}{\sqrt{2\pi r}}C(\omega)e^{iq_{\rm pl}r}T(q_{\rm pl} \hat{\bm r},\theta,\omega)~.
\end{eqnarray}
In the second approximate equality we used the asymptotic expression in 
Eq.(\ref{eq:effectiveinteractionbehaviour}). Comparing Eq.~(\ref{eq:asymptoticbehaviour}) in the main text 
with the result of Eq.~(\ref{eq:lipmannscwingerasymptotic}), we finally obtain Eq.~(\ref{eq:scatteringamplitude}).

\subsection{Geometry in Fig.~\ref{fig:geometry}(b)}
The Lippmann-Schwinger equation for this geometry is
\begin{equation}\label{eq:LS_1b_geometry}
\begin{split}
& V_{\rm sc}(x,\theta,\omega) = V^{(0)}(x,\theta,\omega)\\
&+\int d x'W_{\rm h}^{(+)}(x-x',q_{\rm pl}\sin(\theta),\omega)\\
& \times \frac{1}{L_x}\int d x'' \Delta(x',x'',q_{\rm pl}\sin(\theta),\omega)V_{\rm sc}(x'',\theta,\omega)~,
\end{split}
\end{equation}
where
\begin{equation}
W^{(\pm)}_{\rm h}\left(x,q_y ,\omega\right)=\int \frac{dq_x}{2\pi}e^{i q_x x}W^{(\pm )}_{\rm h}(q,\omega)~,
\end{equation}
with $q$ defined as right after Eq.~(\ref{eq:inversepotential}), and 
\begin{widetext}
\begin{equation}
\begin{split}
W^{(\pm )}_{\rm h}(\sqrt{q_{\rm pl}^2 \sin^2(\theta)+q_x^2},\omega) & = \frac{C(\omega)}{\cos(\theta)}\left[{\cal P}\frac{1}{q_{\rm pl}\cos(\theta)+q_x} + {\cal P}\frac{1}{q_{\rm pl}\cos(\theta)-q_x}\mp i\pi \delta(q_{\rm pl}\cos(\theta)-q_x)\mp i\pi \delta(q_{\rm pl}\cos(\theta)+q_x)\right]\\
&+ C(\omega)\left[\frac{\sqrt{q_{\rm pl}^2 \sin^2(\theta)+q_x^2}-q_{\rm pl}}{q_{\rm pl}^2\cos^2(\theta)-q_x^2}
+\frac{1}{\sqrt{q_{\rm pl}^2 \sin^2(\theta)+q_x^2}}+\frac{\mathcal{W}\left(\sqrt{q_{\rm pl}^2 \sin^2(\theta)+q_x^2},\omega\right)}{q_{\rm pl}}
\right]~.
\end{split}
\end{equation}
\end{widetext}
The asymptotic behavior for $|x|\gg (q_{\rm pl}\cos\theta)^{-1}$ is completely controlled by the divergent terms in the first line. Fourier transforming only this part, we obtain the asymptotic behavior:
\begin{equation}\label{eq:B7}
W^{(\pm)}_{\rm h}\left(x,q_{\rm pl }\sin(\theta) ,\omega\right) \simeq \mp \frac{iC(\omega)}{\cos(\theta)}e^{\pm i q_{\rm pl}\cos(\theta) |x|}~.
\end{equation}

Using Eq.~(\ref{eq:B7}) in Eq.~(\ref{eq:LS_1b_geometry}), we finally find:
\begin{widetext}
\begin{equation}
\begin{split}
V_{\rm sc}(x,\theta,\omega) & 
\simeq \begin{cases}
e^{i q_{\rm pl}\cos\theta x}-\frac{iC(\omega)}{\cos\theta}e^{i q_{\rm pl}\cos\theta x}\int d x' e^{-i q_{\rm pl}\cos\theta x}\frac{1}{L_x} \int d x'' \Delta(x',x'',q_{\rm pl} \sin\theta,\omega)V_{\rm sc}(x'',\theta,\omega),~x\to +\infty\\
e^{i q_{\rm pl}\cos\theta x}-\frac{iC(\omega)}{\cos\theta}e^{-i q_{\rm pl}\cos\theta x}\int d x' e^{+i q_{\rm pl}\cos\theta x}\frac{1}{L_x}\int d x'' \Delta(x',x'',q_{\rm pl} \sin\theta,\omega)V_{\rm sc}(x'',\theta,\omega),~x\to -\infty\\
\end{cases}\\
& \simeq \begin{cases}
e^{i q_{\rm pl}\cos\theta x}\left(1-\frac{iC(\omega)}{\cos\theta} \right)T(q_{\rm pl}\cos\theta,\theta,\omega),~x\to +\infty\\
e^{i q_{\rm pl}\cos\theta x}-\frac{iC(\omega)}{\cos\theta}T(-q_{\rm pl}\cos\theta,\theta,\omega)e^{-i q_{\rm pl}\cos\theta x},~x\to -\infty~.
\end{cases}
\end{split}
\end{equation}
\end{widetext}
Comparing this result with Eq.~(\ref{eq:asymptoticbehaviour1d}) in the main text, we obtain the desired expressions for the transmission and reflection coefficients listed in Eqs.~(\ref{eq:transcoeff}) and~(\ref{eq:refcoeff}).
\section{Going beyond the RPA}
\label{app:manybody}

The simplest way of transcending the RPA~\cite{Giuliani_and_Vignale} 
(i.e.~the time-dependent Hartree approximation discussed in Sect.~\ref{sect:chi}) consists in using time-dependent density-functional theory 
(TDDFT)~\cite{Giuliani_and_Vignale,ullrich_prb_2002}. This theory is appealing since, as we proceed to show, it requires very little modifications of our scattering equations.

We define the Kohn-Sham response function~\cite{Giuliani_and_Vignale}
\begin{equation}
\chi_{nn}^{\rm KS}(\bm q,\bm q',\omega)\equiv \frac{1}{S}\sum_{\alpha, \beta} \frac{(f_\alpha-f_\beta)\langle \alpha| n_{\bm q}|\beta\rangle\langle \beta| n_{-\bm q'}|\alpha\rangle}{\hbar \omega +\epsilon_\alpha-\epsilon_\beta+i\eta}~,
\end{equation}
where $|\alpha\rangle$, $\epsilon_\alpha$ are eigenstates and eigenvalues of the self-consistent Kohn-Sham Hamiltonian~\cite{Giuliani_and_Vignale}. The density perturbation generated by an external field is given by~\cite{Giuliani_and_Vignale,ullrich_prb_2002}
\begin{eqnarray}\label{eq:induceddensityKS}
n_{1}(\bm q,\omega) & = & 
\sum_{\bm q'} \chi^{\rm KS}_{nn}(\bm q,\bm q',\omega)\Big\{V_{\rm ext}(\bm q',\omega)\nonumber\\
& + & \sum_{\bm q''}\left[v(\bm q',\bm q'',\omega) + f_{\rm xc, L}(\bm q',\bm q'',\omega)\right]n_1(\bm q'',\omega)\Big\}~.\nonumber\\
\end{eqnarray}
The first term in curly brackets is the response of the non-interacting Kohn-Sham electron system, while the second and third terms stem from the time variation of the Hartree and exchange-correlation potentials, respectively. The quantity $f_{\rm xc, L}(\bm q',\bm q'',\omega)$ is the wavevector- and frequency-dependent (longitudinal) exchange-correlation kernel~\cite{Giuliani_and_Vignale}. We now introduce---cf. Eqs.~(\ref{eq:chisplitting}) and~(\ref{eq:potentialsplitting})---the following decompositions:
\begin{equation}\label{eq:decomposition_chi_Kohn_Sham}
\chi_{nn}^{\rm KS}(\bm q,\bm q',\omega)=\delta_{\bm q, \bm q'}\chi_0(q,\omega)+\frac{1}{S}\delta \chi_{nn}^{\rm KS}(\bm q,\bm q',\omega)~,
\end{equation}
and
\begin{equation}\label{eq:decomposition_fxc}
f_{\rm xc, L}(\bm q,\bm q',\omega)=\delta_{\bm q, \bm q'}f_{\rm xc, h}(q, \omega)+\frac{1}{S}\delta f_{\rm xc}(\bm q,\bm q',\omega)~.
\end{equation}
In writing Eq.~(\ref{eq:decomposition_chi_Kohn_Sham}) we used the fact that the homogeneous part of the Kohn-Sham response function coincides with the non-interacting response function $\chi_0(q,\omega)$ of the homogeneous electron system in absence of perturbation.
 
Comparing Eq.~(\ref{eq:induceddensityKS}) with Eqs.~(\ref{eq:induceddensity})-(\ref{eq:densitytopotential}), it is straightforward to show that the TDDFT version of our Lippmann-Schwinger scattering theory can be written down with the following replacements:
\begin{align}
&\tilde{\chi}_{\rm h}(q,\omega) \mapsto \chi_{0}( q,\omega)~,\\
&\delta\tilde{\chi}(\bm q,\bm q',\omega) \mapsto \delta\chi_{nn}^{\rm KS}(\bm q, \bm q',\omega)~,\\
&\epsilon_{\rm h}(q,\omega) \mapsto 1-\left[v(q,\omega)+f_{\rm xc, h}(q,\omega)\right]\chi_{0}(q,\omega)~,\\
&\Delta(\bm q,\bm q',\omega) \mapsto \Delta(\bm q,\bm q',\omega) + \Delta_{\rm xc}(\bm q,\bm q',\omega)~,
\end{align}
where 
\begin{widetext}
\begin{equation}
\begin{split}
\Delta_{\rm xc}(\bm q,\bm q',\omega)=\frac{f_{\rm xc, h}(q,\omega)}{v(q,\omega)}\delta\chi_{nn}^{\rm KS}(\bm q, \bm q',\omega)
+\frac{\delta f_{\rm xc}(\bm q, \bm q',\omega)}{v(q, \omega)}\chi_{0}(q', \omega)
+\frac{1}{S}\sum_{\bm q''}\frac{\delta f_{\rm xc}(\bm q, \bm q'',\omega)}{v(q, \omega)}\delta\chi_{nn}^{\rm KS}(\bm q'',\bm q',\omega)~.
\end{split}
\end{equation}
\end{widetext}
Explicit calculations of the scattering kernel $\Delta_{\rm xc}(\bm q,\bm q',\omega)$ requite explicit expressions for the exchange-correlation kernel $f_{\rm xc, L}(\bm q,\bm q',\omega)$ of the inhomogeneous electron system. To this aim, we refer the reader to Ref.~\onlinecite{Giuliani_and_Vignale} and references therein. 

\section{Static stress tensor response of a two-dimensional electron liquid to a scalar potential}
\label{app:stress}

In this Appendix we calculate the stress tensor of a 2D electron system subject to an external scalar potential.

We consider Eq.~(\ref{eq:stresstensordef}) with ${\bm A}({\bm r}) = {\bm 0}$. The average value of the stress-tensor operator is given, up to linear order in the external field, by
\begin{equation}
\langle T_{\bm q, \alpha \beta} \rangle= \langle T_{\bm q, \alpha \beta} \rangle_0+\sum_{\bm q} \chi_{\alpha\beta}(\bm q) U_{\rm ext}(\bm q)~,
\end{equation}
where $\langle \dots \rangle_{0}$ indicates an average over the ground state of the homogeneous electron liquid, and $\chi_{\alpha\beta}(\bm q)$ is the static density-stress tensor response function. The latter can be expressed using Kubo formula~\cite{Giuliani_and_Vignale}: 
\begin{equation}\label{eq:responsefunctiondef}
\chi_{\alpha\beta}(\bm q)=-\frac{i}{\hbar S}\lim_{\eta \to 0^+}\int_0^\infty d\tau e^{-\eta t} \langle [T_{\bm q, \alpha\beta}(\tau), n_{-\bm q}]\rangle_0~,
\end{equation}
where $T_{\bm q, \alpha\beta}(\tau)$ is the stress tensor operator at time $\tau$ in the Heisenberg representation---see Sect.~\ref{sect:chi}---and the  expectation value must be taken over the ground state of the unperturbed {\it interacting} electron liquid.

In the RPA, we can replace the response function of the interacting electron system with
\begin{equation}
\chi_{\alpha\beta}(\bm q)=\frac{\chi_{\alpha\beta}^{(0)}(\bm q)}{1-v(q, \omega =0)\chi_{0}(q,\omega=0)}~,
\end{equation}
where $\chi_{\alpha\beta}^{(0)}(\bm q)$ is the static density-stress tensor response function of the non-interacting electron system and $\chi_{0}(q,\omega=0)$ is the non-interacting static density-density response function~\cite{Giuliani_and_Vignale}. 

For a non-interacting 2D parabolic-band electron system, the right-hand side of Eq.~(\ref{eq:responsefunctiondef}) can be easily calculated. We find 
\begin{equation}
\begin{split}
\chi_{\alpha\beta}^{(0)}(\bm q) & =\frac{g}{S}\sum_{\bm k}\frac{f_{\bm k}-f_{\bm k+\bm q}}{\epsilon_{\bm k}-\epsilon_{\bm k+\bm q}}\langle \bm k |T_{\bm q, \alpha\beta}|\bm k +\bm q \rangle\\
& =-\frac{2g}{S}\sum_{\bm k}\frac{f_{\bm k}}{\frac{\hbar^2}{2m}\left[q^2+2\bm k \cdot \bm q \right]}\langle \bm k |T_{\bm q, \alpha\beta}|\bm k +\bm q \rangle~,
\end{split}
\end{equation}
where $g=2$ is a spin degeneracy factor, $f_{\bm k} = \Theta(k_{\rm F}-k)$ is the usual zero-temperature Fermi step, and $\epsilon_{\bm k}=\hbar^2 {\bm k}^2/(2m)$ is the band energy.

Making use of the following matrix element of the stress tensor between plane-wave states,
\begin{equation}
\langle \bm k |T_{\bm q, \alpha\beta}|\bm k +\bm q \rangle=E_{\rm F} \left(2\bar{k}_\alpha\bar{k}_\beta+\bar{q}_\alpha \bar{k}_\beta +\bar{k}_\alpha\bar{q}_\beta  + \bar{q}_\alpha\bar{q}_\beta \right)~,
\end{equation}
where $\bar{q}=q/k_{\rm F}$, $\bar{k}=k/k_{\rm F}$, and assuming, without loss of generality, that $\bm q$ lies in the $\hat{\bm x}$ direction, we get:
\begin{widetext}
\begin{equation}
\begin{split}
\chi_{\alpha\beta}^{(0)}(\hat{\bm x} q) & =-\frac{2 E_{\rm F} N_0}{\bar{q}}\int_0^1 d \bar{k} \bar{k} \int_{-\pi}^{\pi}\frac{d \theta}{2 \pi}\frac{2\bar{k}^2\cos^2(\theta)\delta_{\alpha x}\delta_{\beta x}+2\bar{k}^2\sin^2(\theta)\delta_{\alpha y}\delta_{\beta y}+2\bar{q}\bar{k}\cos(\theta)\delta_{\alpha x}\delta_{\beta x}+\bar{q}^2\delta_{\alpha x}\delta_{\beta x}}{\frac{\bar{q}}{2}+\bar{k}\cos(\theta)}\\
& =- E_{\rm F} N_0\frac{(-2)}{\bar{q}}\int_0^1 d \bar{k} \bar{k} \int_{-\pi}^{\pi}\frac{d \theta}{2 \pi}\frac{\delta_{\alpha \beta}\left(\bar{k}^2+\bar{k}\bar{q}\cos(\theta)+\frac{1}{2} \bar{q}^2\right)+\sigma^{(3)}_{\alpha \beta}\left(\bar{k}^2\cos(2\theta)+\bar{k}\bar{q}\cos(\theta)+\frac{1}{2} \bar{q}^2\right)}{-\frac{\bar{q}}{2}-\bar{k}\cos(\theta)}\\
& =-N_{0} E_{\rm F}[f_{\rm s}\left(\bar{q}\right)\delta_{\alpha \beta} + f_{\rm a}\left(\bar{q}\right)\sigma^{(3)}_{\alpha \beta}]~.
\end{split}
\end{equation}
\end{widetext}
Here, the functions $f_{\rm s/a}(\bar{q})$ 
can be expressed in terms of the auxiliary functions $\psi^{(n,\ell)}(z)$ defined by
\begin{equation}\label{eq:psifunction2D}
\psi^{(n,\ell)}(z)\equiv \int_{0}^{1}dx~x^{1 + n + \ell} 
\int_{-\pi}^{\pi} \frac{d\theta}{2\pi} \frac{\cos(\ell\theta)}{z-x\cos(\theta)}~.
\end{equation}
Explicit expressions for these functions are provided in Table \ref{tab:psifunct}.
\begin{table}[t]
\begin{tabular}{ccc}
\hline
$n$ & $\ell$ & $\psi^{(n,\ell)}(z)$ \\
\hline
\hline
$0$ & $0$ & $z-\lambda\sqrt{z^2-1}$\\
$2$ & $0$ & $\frac{2}{3}z^3-\frac{\lambda(1+2z^2)}{3} \sqrt{z^2-1}$\\
$0$ & $1$ & $z^2-\frac{1}{2}-\lambda z\sqrt{z^2-1}$\\
$0$ & $2$ & $\frac{4}{3}z^3-z+\frac{\lambda(1-4z^2)}{3}\sqrt{z^2-1}$\\
\end{tabular}
\caption{\label{tab:psifunct}Explicit expressions for the functions $\psi^{(n,\ell)}(z)$. 
Here $\lambda={\rm sign}[\Re e~(z^*\sqrt{z^2-1})]$, and the function ${\rm sign}(x)$ evaluates to $0$ in $x=0$.}
\end{table}

Putting everything together, we finally obtain
\begin{equation}
\begin{split}
f_{\rm s}(\bar{q}) & =\bar{q} \psi^{(0,0)}\left(\frac{\bar{q}}{2}\right)-2 \psi^{(0,1)}\left(\frac{\bar{q}}{2}\right)+ \frac{2}{\bar{q}}\psi^{(2,0)}\left(\frac{\bar{q}}{2}\right)\\
& = 1+\frac{\bar{q}^2}{6}-\frac{1}{3 |\bar{q}|}\left(\frac{\bar{q}^2}{2}+1\right)\Theta(\bar{q}-2)\sqrt{\bar{q}^2-4}
\end{split}
\end{equation}
and
\begin{equation}
\begin{split}
f_{\rm a}(\bar{q})
& =\bar{q}\psi^{(0,0)}\left(\frac{\bar{q}}{2}\right)-2 \psi^{(0,1)}\left(\frac{\bar{q}}{2}\right)+\frac{2}{\bar{q}} \psi^{(0,2)}\left(\frac{\bar{q}}{2}\right)\\
& = \frac{\bar{q}^2}{3}-\frac{1}{3 |\bar{q}|}\left(\bar{q}^2-1\right)\Theta(\bar{q}-2)\sqrt{\bar{q}^2-4}~.
\end{split}
\end{equation}
The quantities $f_{x/y}(q)$ defined in Eqs.~(\ref{eq:fxdefinition})-(\ref{eq:fydefinition}) are related to $f_{\rm s}$ and $f_{\rm a}$ by $f_{x/y}(q)=[f_{\rm s}(q/k_{\rm F})\pm f_{\rm a}(q/k_{\rm F})]/2$.

\end{document}